\documentstyle[aps,pre,multicol]{revtex}
\input psfig

\begin{document}

\title{Dynamics of simulated water under pressure}

\author{Francis~W. Starr$^{1}$, Francesco Sciortino$^{2}$, and H. Eugene
Stanley$^{1}$}

\address{$^1$Center for Polymer Studies, Center for Computational
Science, and Department of Physics, \linebreak Boston University,
Boston, MA 02215 USA}

\address{$^2$Dipartmento di Fisica e Istituto Nazionale per la Fisica
della Materia,\\ Universit\'{a} di Roma ``La Sapienza'', Piazzale Aldo
Moro 2, I-00185, Roma, Italy}

\date{August 2, 1999; Submitted to {\it Physical Review E}}

\maketitle

\begin{abstract}
We present molecular dynamics simulations of the SPC/E model of water to
probe the dynamic properties at temperatures from 350~K down to 190~K
and pressures from 2.5~GPa (25~kbar) down to -300~MPa ($-3$~kbar).  We
compare our results with those obtained experimentally, both of which
show a diffusivity maximum as a function of pressure.  We find that our
simulation results are consistent with the predictions of the
mode-coupling theory (MCT) for the dynamics of weakly supercooled
liquids -- strongly supporting the hypothesis that the apparent
divergences of {\it dynamic\/} properties observed experimentally may be
independent of a possible thermodynamic singularity at low temperature.
The dramatic change in water's dynamic and structural properties as a
function of pressure allows us to confirm the predictions of MCT over a
much broader range of the von Schweidler exponent values than has been
studied for simple atomic liquids.  We also show how structural changes
are reflected in the wave-vector dependence of dynamic properties of the
liquid along a path of nearly constant diffusivity.  For temperatures
below the crossover temperature of MCT (where the predictions of MCT are
expected to fail), we find tentative evidence for a crossover of the
temperature dependence of the diffusivity from power-law to Arrhenius
behavior, with an activation energy typical of a strong liquid.
\end{abstract}
\bigskip

\begin{multicols}{2}

\section{Introduction}

The ``slow dynamics'' and glass transition of both simple and molecular
liquids has been a topic of significant interest in recent years.  The
initial slowing down of liquids at temperatures down to $T_c \approx
1.2T_g$, where relaxation times approach $1$~ns, has been well described
by the mode-coupling theory (MCT)~\cite{mct}.  MCT has been successfully
applied to a wide variety of real and model systems~\cite{mctbook},
including hard spheres~\cite{hardspheres},
Ni$_{80}$P$_{20}$~\cite{ka95}, SiO$_2$~\cite{silica}, and polymer
melts~\cite{benn}.  However, there has not been an extensive test of the
validity of the MCT predictions for a model system over a wide range of
pressures and along different thermodynamic paths.

At low pressure, it was shown previously that the power-law behavior of
dynamic properties in the SPC/E model~\cite{spce} can be explained using
MCT~\cite{francesco}.  Furthermore, the possible relationship between
the experimentally-observed power-law behavior and the predictions of
MCT has been discussed~\cite{francesco,prielmeier,germans,xenon}.  The
experimentally-observed locus of apparent power-law singularities of
dynamic and thermodynamic properties [Fig.~\ref{fig:water-T_s}] is of
particular interest~\cite{angell81,debenedetti}, and has catalysed the
development of three scenarios to explain the anomalous properties of
water: (i) the existence of a spinodal bounding the stability of the
liquid in the superheated, stretched, and supercooled
states~\cite{angell81,spinodal}; (ii) the existence of a liquid-liquid
phase transition line separating two liquid phases differing in
density~\cite{pses,mishima94,critical-point,ms98}; (iii) a
singularity-free scenario in which the thermodynamic anomalies are
related to the presence of low-density and low-entropy structural
heterogeneities~\cite{singfree}.  The predictions of MCT are of interest
since MCT might account for the apparent power-law behavior of dynamic
properties on cooling, thereby removing the need for a thermodynamic
explanation of the dynamic properties of water.

In this article we focus on two related issues: (i) the possibility of
using MCT to explain the slow dynamics of water under pressure, and (ii)
a test of the validity of the MCT predictions over an extremely wide
pressure range in a system with dramatic structural changes.  We find
that MCT provides a good account of the slow dynamics of the SPC/E model
for water at all pressures, with the structure evolving continuously
from an open tetrahedral network to a densely packed fluid, similar to a
Lennard-Jones type liquid.  By examining the wave-vector dependence of
collective dynamics, we are able to discover how these structural
changes are reflected in the dynamic behavior of the liquid.  We are
also able to test the validity of the relationship predicted by MCT for
the diffusivity exponent $\gamma$ and the von Schweidler exponent $b$
over a wide range of values $\gamma$ and $b$ [Fig.~\ref{fig:b-gamma}].
Our results support the predicted relationship of these exponents. A
brief report of a subset of the present results for the SPC/E potential
has recently appeared~\cite{shss}.  The dynamic properties of the ST2
potential~\cite{paschek-geiger} at one pressure, and of the
TIP4P~\cite{tanaka} potential at several pressures, have also recently
been discussed.

\newbox\figa
\setbox\figa=\psfig{figure=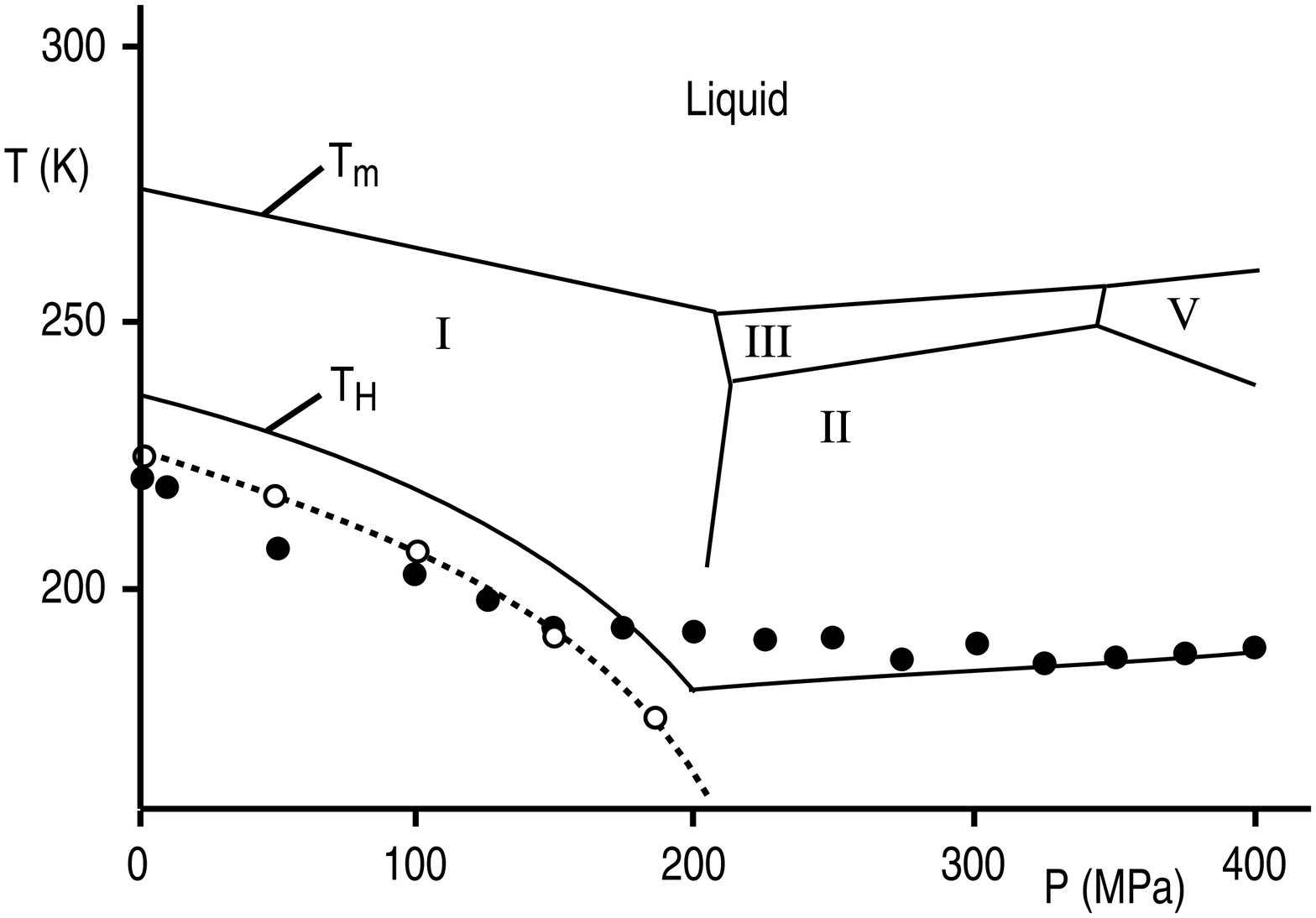,width=3.35in}
\begin{figure*}[htbp]
\begin{center}
\leavevmode
\centerline{\box\figa}
\narrowtext
\caption{Phase diagram of water.  The extrapolated divergence of the
isothermal compressibility ($\circ$)~\protect\cite{kanno-angell79} and
the extrapolated divergence of $D$
($\bullet$)~\protect\cite{prielmeier}.  The different loci of these two
singularity lines are consistent with the possibility that the two
phenomena may arise from different explanations.  Also shown are the
melting line (T$_m$) and coexistence lines of several ice polymorphs and
the experimental limit of supercooling (T$_H$). }
\label{fig:water-T_s}
\end{center}
\end{figure*}

\section{Mode Coupling Theory}

We will focus our discussion on the idealized form of MCT, originally
formulated to describe spherically-symmetric potentials.  Recent
extensions have been made to account for the rotational motion present
in non-spherical molecular systems~\cite{molecular-mct}, such as water.
The idealized version of MCT has been shown to provide a good account
for the center-of-mass motion for the SPC/E
model~\cite{francesco,not-com}.  We provide only a brief account of the
MCT predictions relevant to the results of this article, and we refer
the reader to extensive reviews for more
information~\cite{mct,mctbook,leshouches}.

MCT assumes that localization, or ``caging'' of molecules due to the
slow rearrangement of neighboring molecules, is the source of the
dramatic increase of relaxation times on cooling, leading to a strong
coupling between single particle motion and the density fluctuations of
the liquid.  Indeed, according to MCT, the {\it static} density
fluctuations, measured by the structure factor $S(q)$, entirely
determine the long time dynamic behavior.  MCT accounts for the loss of
correlation by the interaction of density mode fluctuations, ignoring
other possible mechanisms for relaxation. MCT predicts the asymptotic
power-law divergence of correlation times, and power-law vanishing of
the diffusion constant
\begin{equation}
D \sim D_0 (T/T_c-1)^\gamma
\label{eq:D-plaw}
\end{equation}
at a critical temperature $T_c = T_c(P)$, where we refer to
$\gamma=\gamma(P)$ as the diffusivity exponent.  In real systems,
``freezing'' of the system dynamics is avoided at $T_c$, as \linebreak 

\newbox\figa
\setbox\figa=\psfig{figure=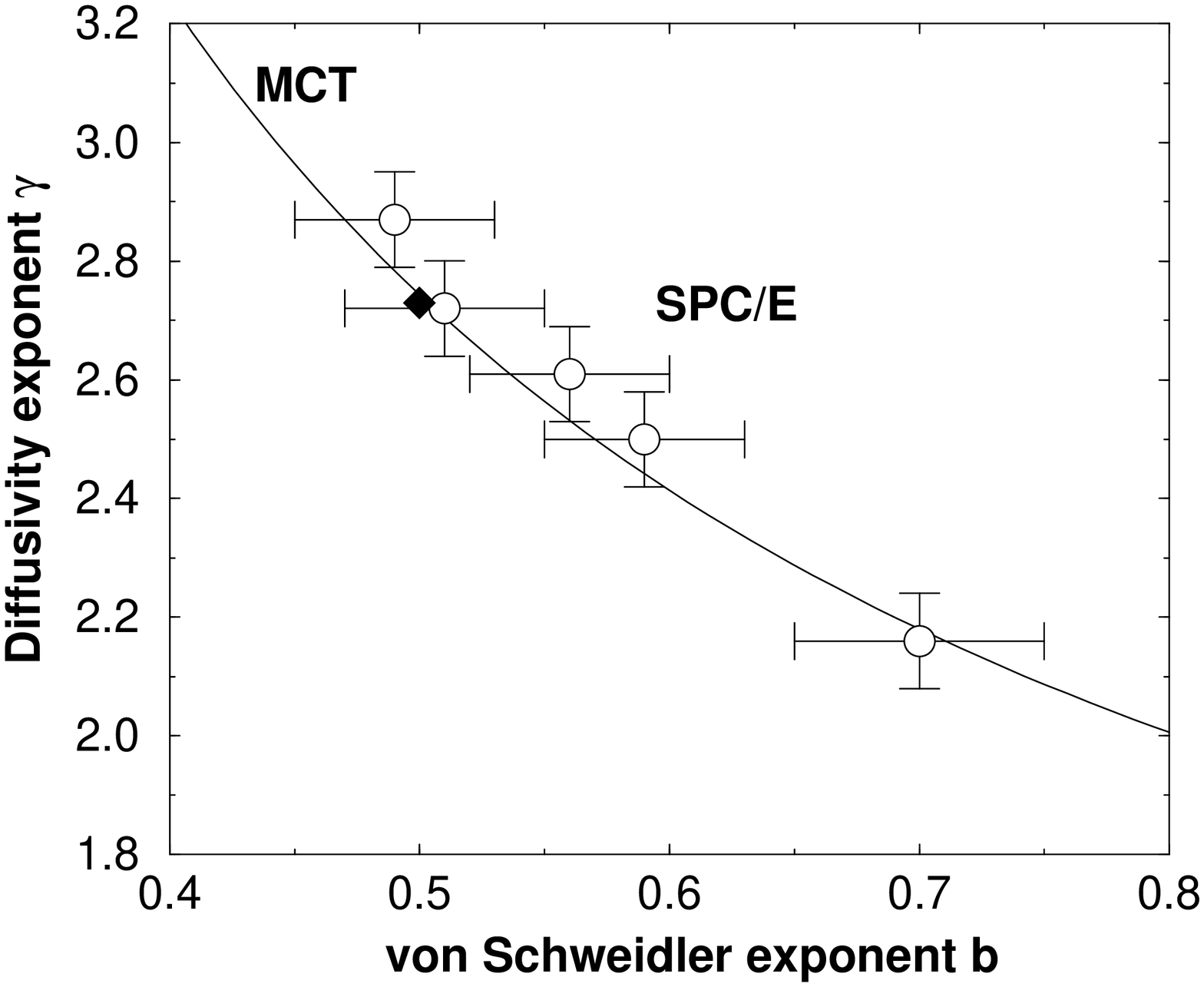,width=3.35in,angle=-90}
\begin{figure*}[htbp]
\begin{center}
\leavevmode
\centerline{\box\figa}
\narrowtext
\caption{The line shows the predicted relationship between $b$ and
$\gamma$ from MCT. The symbols show the calculated values for the SPC/E
model: ($\circ$) from this work, (filled $\Diamond$) from
Ref.~\protect\cite{francesco}.}
\label{fig:b-gamma}
\end{center}
\end{figure*}
\noindent relaxation mechanisms not accounted for by MCT become
significant.  However, $T_c$ can still be interpreted as a ``crossover
temperature'' where the dynamics change from being dominated by density
fluctuations to being controlled by ``activated'' processes.  Some
recent work has also demonstrated the significance of $T_c$ as a
crossover temperature where relaxation occurs primarily through basin
hopping~\cite{st98,dynamic-het,otp,schroder}, in the energy landscape
view of liquid dynamics~\cite{goldstein,glasses95}.

MCT predicts that the Fourier transform of the density-density
correlation function~\cite{hansen-mcdonald} or intermediate scattering
function 
\begin{equation}
F(q,t) \equiv \frac{1}{S(q)} \left\langle \sum_{j,k=1}^N e^{-i {\bf
q}\cdot[{\bf r}_k(t) - {\bf r}_j(0)]} \right\rangle
\label{eq:isf}
\end{equation}
decays via a two-step process.  In the first relaxation step, $F(q,t)$
approaches a plateau value $F_{\mbox{\scriptsize plateau}}(q)$ which is
described, to leading order in time, by a power law with exponent $a =
a(P)$,
\begin{equation}
F(q,t) - F_{\mbox{\scriptsize plateau}}(q) \sim t^a,
\label{eq:plateau-approach}
\end{equation}
At larger times, $F(q,t)$ decreases from $F_{\mbox{\scriptsize
plateau}}(q)$ and MCT predicts the decay obeys the von Schweidler power
law to leading order in time
\begin{equation}
F_{\mbox{\scriptsize plateau}}(q) - F(q,t) \sim t^b,
\label{eq:vonSchweidler}
\end{equation}
where $b=b(P)$ is known as the von Schweidler exponent.  The region of
validity of Eqs.~(\ref{eq:plateau-approach}) and
(\ref{eq:vonSchweidler}) can be quite limited.

The slow relaxation of $F(q,t)$ has a characteristic relaxation time
$\tau$ that is also predicted to have asymptotic power law dependence on
temperature,
\begin{equation}
\tau \sim \tau_0 (T/T_c-1)^{-\gamma}
\label{eq:tau-plaw}
\end{equation}
with the same value of the exponent $\gamma$ as for the diffusion
constant. Hence, Eqs.~(\ref{eq:D-plaw}) and (\ref{eq:tau-plaw}) predict
that the product $D\tau$ is not singular as $T\rightarrow T_c$, hence we
take the product to be constant over the range that
Eqs.~(\ref{eq:D-plaw}) and (\ref{eq:tau-plaw}) are valid (neglecting
corrections to scaling).

MCT predicts that the scaling exponents $a$, $b$, and $\gamma$ are {\it
not} independent; $a$ and $b$ are related by the exponent parameter
$\lambda$ using the relationship
\begin{equation}
\lambda = \frac{[\Gamma(1-a)]^2}{\Gamma(1-2a)} =
\frac{[\Gamma(1-b)]^2}{\Gamma(1+2b)}
\label{eq:abg1}
\end{equation}
where $\Gamma(x)$ is the gamma function.  MCT also relates $\gamma$ to
$a$ and $b$ via
\begin{equation}
\gamma = \frac{1}{2a} + \frac{1}{2b}.
\label{eq:abg2}
\end{equation}
Because of Eqs.~(\ref{eq:abg1}) and (\ref{eq:abg2}), only one exponent
value is needed to determine all others, so calculation of two exponents
determines if the dynamics of a system are consistent with the
predictions of MCT.  Furthermore, these exponents are expected to depend
on the path along which $T_c$ is approached.

After $F(q,t)$ departs from the plateau, $F(q,t)$ is well-described by a
Kohlrausch-Williams-Watts stretched exponential
\begin{equation}
F(q,t) = A(q)
\exp{\left[-\left(\frac{t}{\tau(q)}\right)^{\beta(q)}\right]},
\label{eq:kww}
\end{equation}
where $\tau(q)$ is the relevant relaxation time.  Moreover, it has been
shown that the exponent $\beta = \beta(q)$ is related to the von
Schweidler exponent~\cite{fuchs-beta}
\begin{equation}
\lim_{q\to\infty} \beta(q) = b.
\label{eq:betaq-limit}
\end{equation}
This relation facilitates evaluation of $b$, since the region of validity
of Eq.~(\ref{eq:vonSchweidler}) is difficult to identify in practice.

\section{Simulations}
\label{sec:simulations}
We perform MD simulations of 216 water molecules interacting via the
SPC/E pair potential~\cite{spce}.  The SPC/E model treats water as a
rigid molecule consisting of three point charges located at the atomic
centers of the oxygen and hydrogen, which have an OH distance of
1.0~\AA\ and HOH angle of 109.47$^\circ$, the tetrahedral angle.  Each
hydrogen has charge $q_H = 0.4238e$, where $e$ is the fundamental unit
of charge, and the oxygen has charge $q_O = -2q_H$.  In addition, the
oxygen atoms of separate molecules interact via a Lennard-Jones
potential with parameters $\sigma = 3.166$ \AA\ and $\epsilon = 0.6502$
kJ/mol.

Our simulation results are summarized in Table~\ref{table:state-points}.
For $T \le 300$~K, we simulate two independent systems to improve
statistics, as the long relaxation time makes time averaging more
difficult.  We equilibrate all simulated state points to constant $T$
and $\rho$ by monitoring the pressure and internal energy.  We control
the temperature using the Berendsen method of rescaling the
velocities~\cite{ber-84}, while the reaction field technique with a
cutoff of 0.79 nm~\cite{steinhauser} accounts for the long-range
Coulombic interactions.  The equations of motion evolve using the SHAKE
algorithm~\cite{rcb-77} with a time step of 1~fs, except at $T=190$,
where a time step of 2~fs is used due to the extremely slow motion of
the molecules.  Equilibration times at high temperatures are relatively
small.  At low $T$, extremely long equilibration times are needed.  The
structural and thermodynamics properties may be obtained after
relatively short equilibration times.  However, dynamic properties show
significant aging effects (i.e. dependence of measured properties on the
chosen starting time) if great care is not taken in equilibration.

For production runs, it is desirable to make measurements in the
isoenergetic/isochoric ensemble (NVE).  However, a small energy drift is
unavoidable for the long runs presented here, so we again employ the
heat bath of Berendsen, using a relaxation time of
200ps~\cite{francesco}.  The large relaxation time prevents an energy
drift but achieves results that are very close to those that would be
found if it were possible to perform a simulation in the NVE ensemble.

Since we perform long runs for many state points, we store the molecular
trajectories $\{ {\bf r}_i,{\bf p}_i \}$ at logarithmic intervals to
avoid storage problems that linear sampling presents.  Specifically, we
sample configurations at times growing in powers of $2$ up a maximum
time $t_{\mbox{\scriptsize max}}$.  We begin a new sampling cycle each
time $t_{\mbox{\scriptsize max}}$ (relative to the cycle starting time)
is reached.  This sampling method allows for calculation of dynamic
properties on time scales spanning eight orders of magnitude (from 1~fs
to 100~ns) using a relatively small amount of disk space. Still, more
then 2 GB of storage was required for storing configurations at
$T=210~K$.  Our simulations have a speed of approximately 200 $\mu$s per
update per molecule on a MIPS R10000 processor, representing a total
calculation of approximately 8.4 years of CPU time, including the
systems of 1728 molecules discussed in the appendix. For the larger
systems, we utilize a parallelized version of our simulation code on
eight processors to improve performance.

\section{Static Structure Factor}

We first summarize the structural properties of our simulations in order
to better understand the relationship between the changes in structure
with the changes in dynamic behavior, which we will detail in
Sec.~\ref{sec:isochrone-dynamics}. Other studies have considered the
structural and thermodynamic properties of SPC/E in a large region of the
$(P.T)$ plane~\cite{bc94,hpss97,bagchi,sbs}, so the present discussion
is brief.

\pagebreak

\newbox\figa
\newbox\figb
\setbox\figa=\psfig{figure=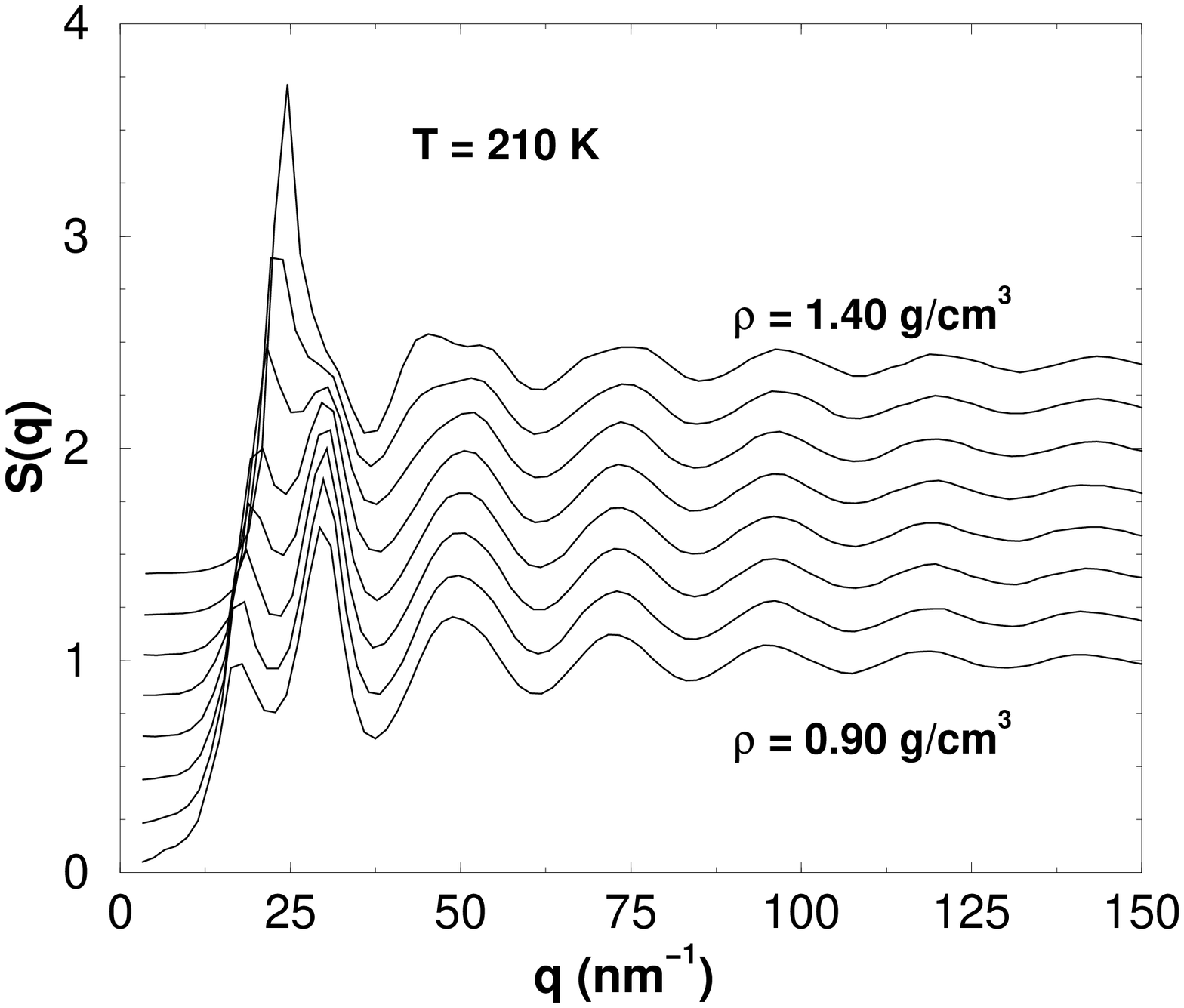,width=3.35in,angle=-90}
\setbox\figb=\psfig{figure=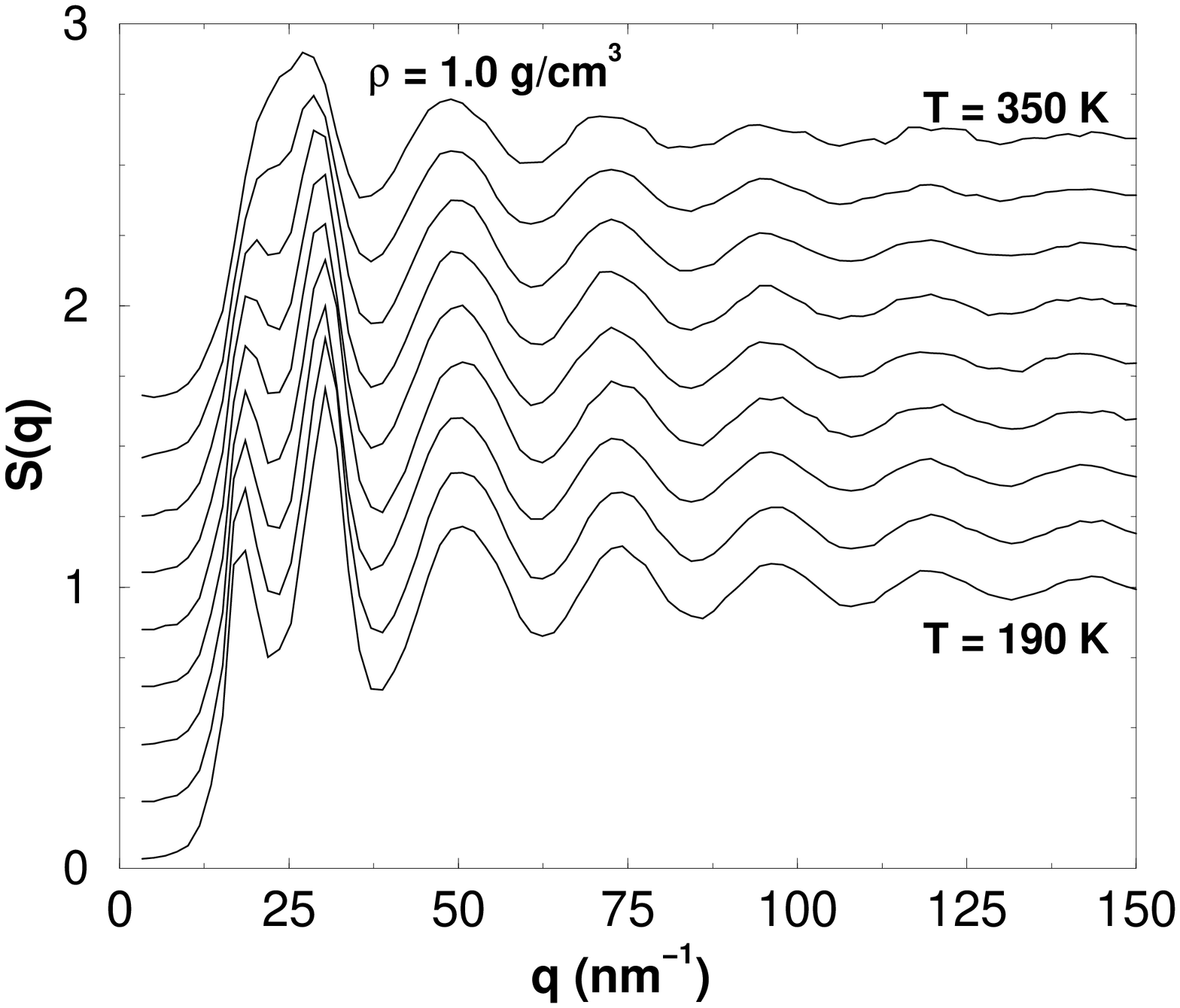,width=3.35in,angle=-90}
\begin{figure*}[htbp]
\begin{center}
\leavevmode
\centerline{\box\figa}
\centerline{\box\figb}
\narrowtext
\caption{The oxygen-oxygen structure factor $S(q)$: (a) Dependence on
$\rho$ for $T=210$~K.  (b) $T$-dependence along the $\rho =
1.0$~g/cm$^3$ isochore.  Notice that changing $T$ has little effect,
while changing $\rho$ has a more pronounced effect.}
\label{fig:sq}
\end{center}
\end{figure*}

The MCT theory requires as input the static density-density correlation
functions. In the case of water, the structure of the system is very
sensitive to the value of the external control parameter $(P,T)$. Hence,
for all state points simulated, we calculate the oxygen-oxygen partial
structure factor~\cite{hansen-mcdonald}
\begin{equation}
S(q)\equiv\frac{1}{N} \left| \sum_{j=1}^N e^{-i {\bf q} \cdot {\bf r}_j}
\right|^2.
\end{equation}
Several studies have carefully calculated the structure of simulated
water, and found surprisingly good agreement with
experiments~\cite{pses,sbs}.  At $T=210$~K, we show the structural
changes from low to high density [Fig.~\ref{fig:sq}(a)].  The structure
at low density/pressure is similar to that observed for low-density
amorphous (LDA) solid water, consisting of an open tetrahedral network.
At high density/pressure, water is very similar to high density
amorphous (HDA) solid water, where core-repulsion dominates, similar to
simple liquids under pressure.

We show the evolution of $S(q)$ as a function of $T$ along the $\rho =
1.0$~g/cm$^3$ isochore in Fig.~\ref{fig:sq}(b).  We note that in the
temperature range from 190~K to 300~K, where the dynamics show the most
dramatic change in behavior, $S(q)$ shows only small changes in the
first two peaks.  Also, the location of the first maximum $q_0$ in
$S(q)$, the wave vector at which $F(q,t)$ typically shows the slowest
relaxation, does not appear to change significantly.  All other
densities and temperatures show a relative smooth interpolation of
Figs.~\ref{fig:sq}(a) and (b).

\section{Mean-Squared Displacement and Diffusion}
\label{sec:diffusion}

The mean-squared displacement $\langle r^2(t) \rangle \equiv \langle |
{\bf r}(t) - {\bf r}(0) |^2 \rangle$ is shown in Fig.~\ref{fig:msd}.
All the curves show $t^2$ dependence at small time, as expected in the
``ballistic'' regime.  For low T (e.g. $T=210$~K
[Fig.~\ref{fig:msd}(a)]), $\langle r^2(t) \rangle$ shows relatively flat
behavior over 3-4 decades in time.  This is the ``cage'' region, in
which a molecule is trapped by its neighbors and cannot diffuse, and is
only vibrating within its ``cage.''  At low $P$, the cage consists of
hydrogen-bonded neighbors in a tetrahedral configuration.  This cage is
relatively strong, compared to simple liquids, because of the H bonds.
The size of the cage may be estimated by the value $\langle r^2(t)
\rangle$ at the plateau, as shown in the inset of
[Fig.~\ref{fig:msd}(a)].  Surprisingly, the size of the cage is not
monotonic with density, and has a maximum at $\rho \approx
1.1$~g/cm$^3$.  We shall see that this corresponds roughly to the $\rho$
at which $D$ also has a maximum.  We observe a small bump in $\langle
r^2(t) \rangle$ at $t \approx 0.35$~ps, as observed in
Ref.~\cite{francesco}.  A system size study indicates that this may be
attributed to finite size effects~\cite{francesco-unpub}.

For long times, all the curves show linear $t$ dependence, indicating
that our simulations are in the diffusive regime.  We extract the
diffusion constant $D$ using the asymptotic relation $\langle r^2(t)
\rangle = 6Dt$.  We plot the density dependence of $D$ in
Fig.~\ref{fig:D-isotherms} and find that the SPC/E potential, like
water, shows an anomalous increase in $D$ on increasing density.  We
also point out the feature that $D$ shows a slight increase at very low
density; namely, at $\rho = 0.90$~g/cm$^3$ and $T = 210$~K.  This can be
attributed to the fact that the liquid is extremely stretched at this
density, causing an increase in the defects of the bond network, and
thus increased diffusivity.  We show $D$ as a function of pressure along
several isotherms to compare with experimental measurements
[Fig.~\ref{fig:DofP}]~\cite{prielmeier}.  The anomalous increase in $D$
is qualitatively reproduced by our calculations for the SPC/E model, but
the quantitative increase of $D$ is significantly larger than
that\linebreak

\newbox\figa
\newbox\figb
\setbox\figa=\psfig{figure=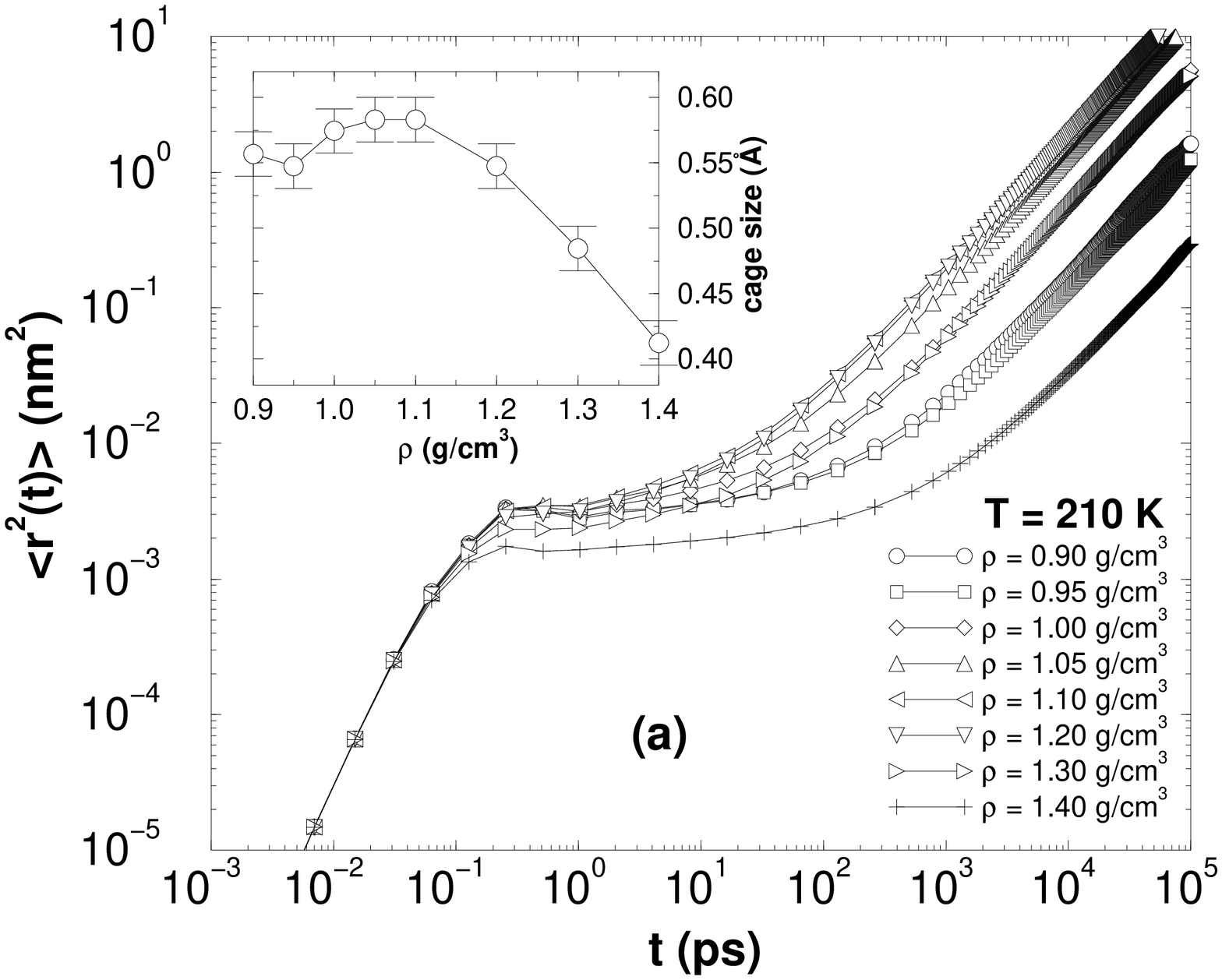,width=3.35in,angle=-90}
\setbox\figb=\psfig{figure=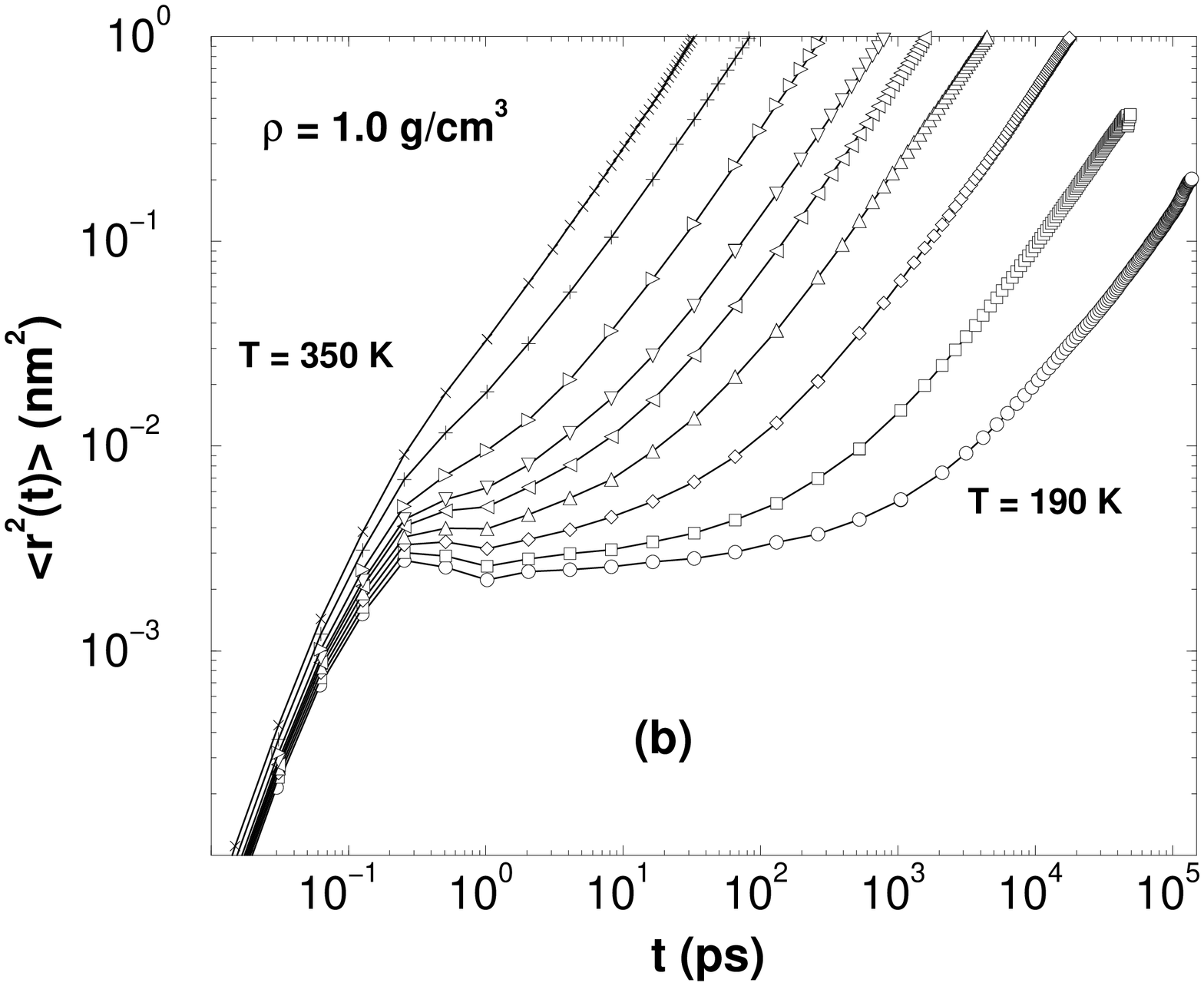,width=3.35in,angle=-90}
\begin{figure*}[htbp]
\begin{center}
\leavevmode
\centerline{\box\figa}
\centerline{\box\figb}
\narrowtext
\caption{Mean-squared displacement $\langle r^2(t) \rangle$ for (a) all
densities at $T=210$~K and (b) all $T$ along the $\rho = 1.0~$g/cm$^3$
isochore.  The inset of (a) shows the density dependence of the cage
size.}
\label{fig:msd}
\end{center}
\end{figure*}
\noindent observed experimentally.  This discrepancy may arise from the
fact that the SPC/E potential is {\it under-structured} relative to
water~\cite{hpss97}, so applying pressure allows for more bond breaking
and thus greater diffusivity than observed experimentally.  We also find
that the pressure where $D$ begins to decrease with pressure --- normal
behavior for a liquid --- is larger than that observed experimentally
\cite{prielmeier}.  This comparison of $D$ with experiment leads us to
expect that while the qualitative dynamic features we observe in the
SPC/E potential may aid in the understanding of the dynamics of water
under pressure, they will likely not be quantitatively accurate.

We estimate $D$ along the isobars $P=-80$~MPa, 0~MPa, 100~MPa, 200~MPa,
300~MPa, and 400~MPa from the isochoric data.  We confirm that along the
-80~MPa isobar, our estimates agree with the $-80$~MPa calculations of
Ref.~\cite{francesco}, which employs the same truncation of the
potential used here (see Sec.~\ref{sec:simulations}).  Along the
0~MPa\linebreak
\newbox\figa
\setbox\figa=\psfig{figure=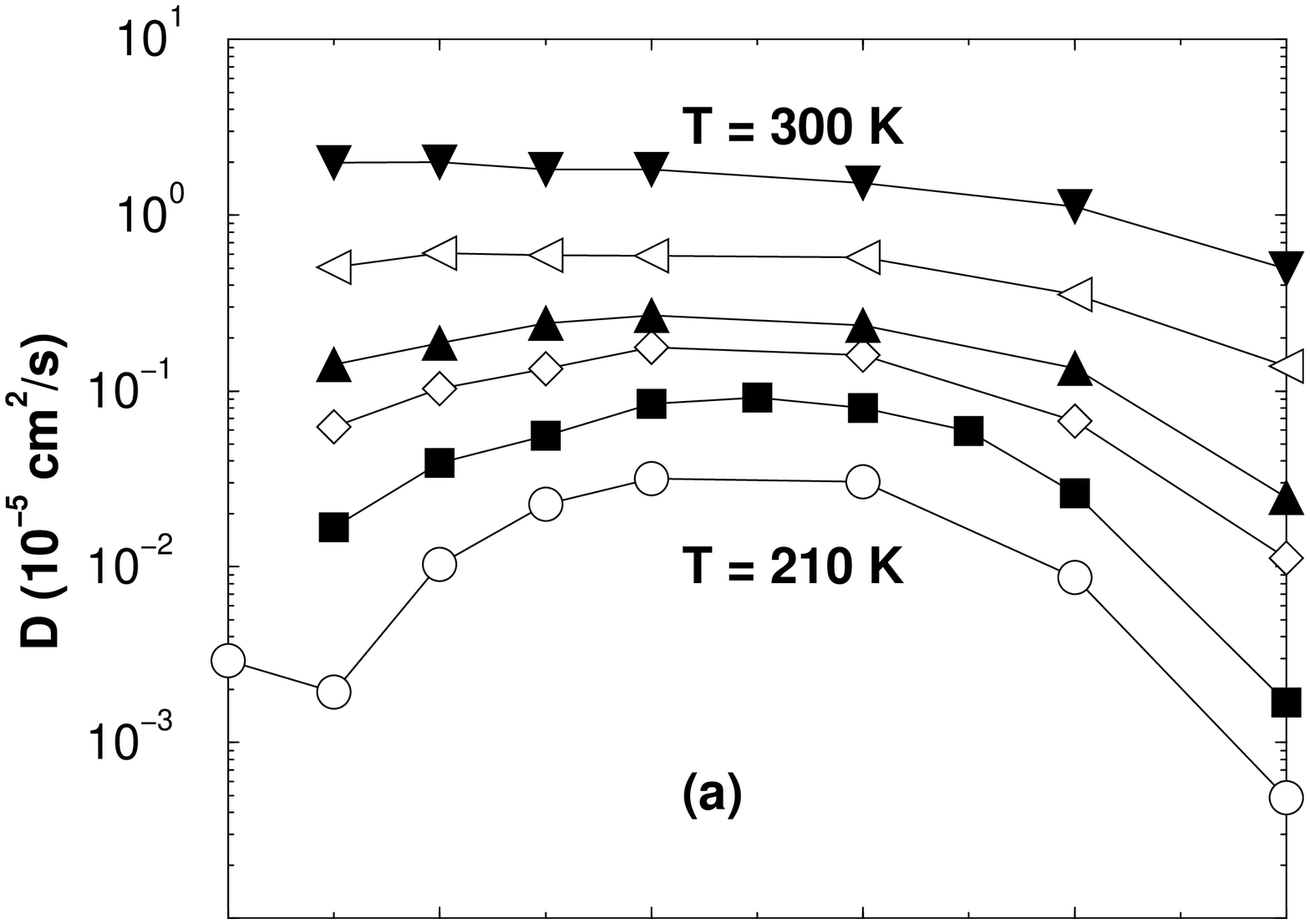,width=3.35in,angle=-90}
\newbox\figb
\setbox\figb=\psfig{figure=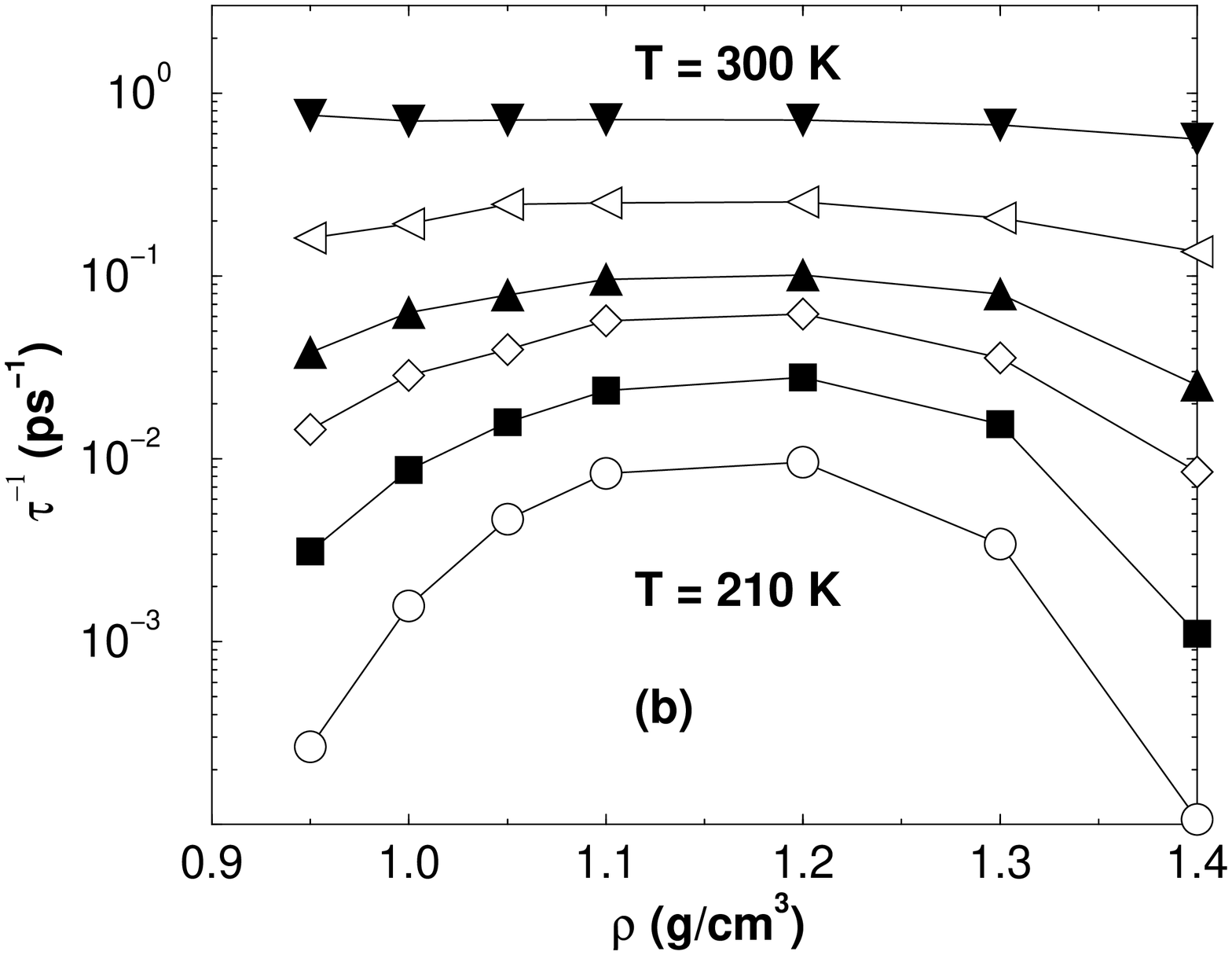,width=3.35in,angle=-90}
\newbox\figc
\setbox\figc=\psfig{figure=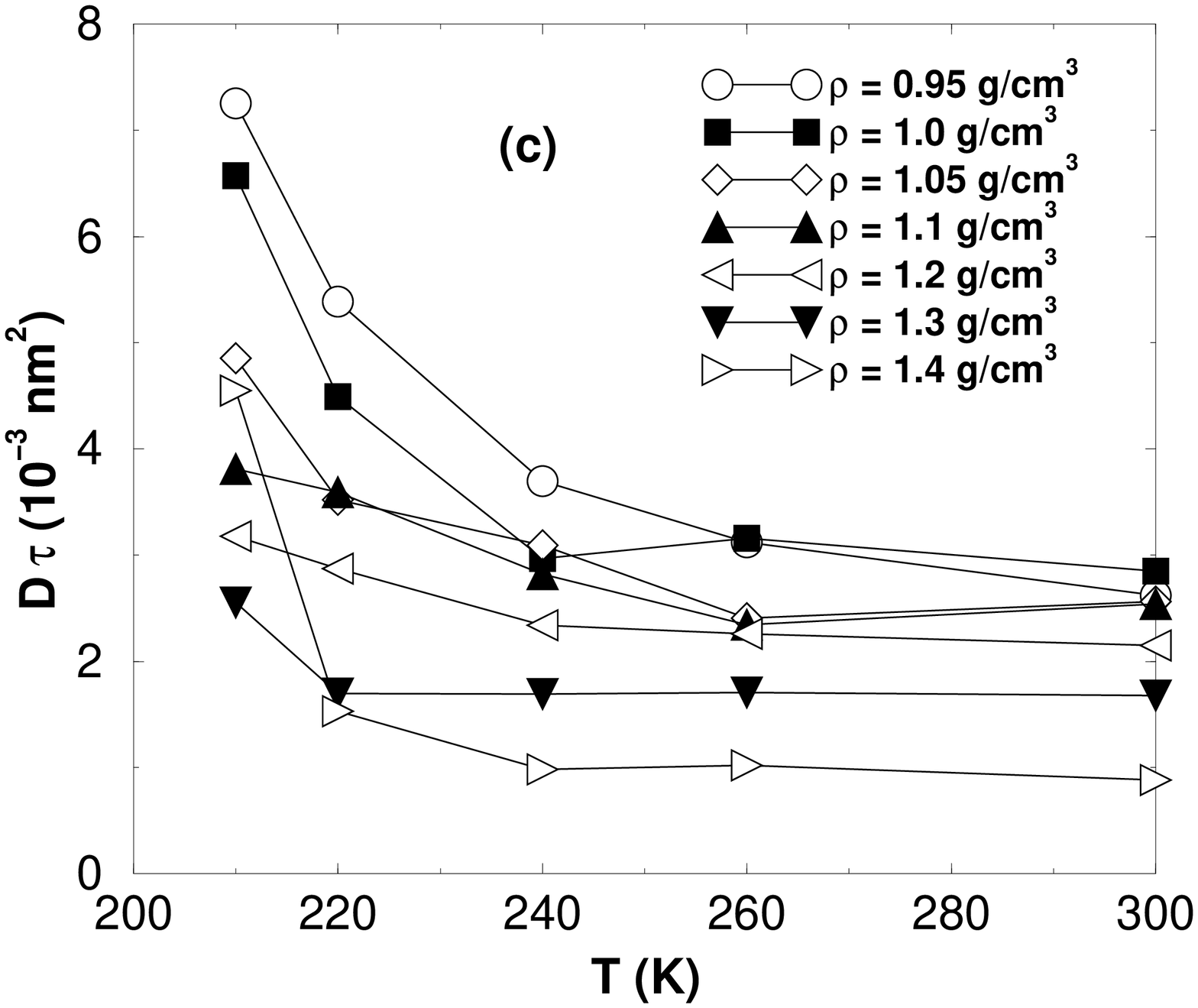,width=3.35in,angle=-90}
\begin{figure*}[htbp]
\begin{center}
\leavevmode
\centerline{\box\figa}
\centerline{\box\figb}
\centerline{\box\figc}
\narrowtext
\caption{(a) Diffusion constant $D$ along isotherms for each density
simulated.  (b) Relaxation time $\tau$ of $F(q_0,t)$ along isotherms
for each density simulated. (c) Test of the MCT prediction that $D \tau$
is constant along isochores.}
\label{fig:D-isotherms}
\end{center}
\end{figure*}
\pagebreak
\newbox\figa
\setbox\figa=\psfig{figure=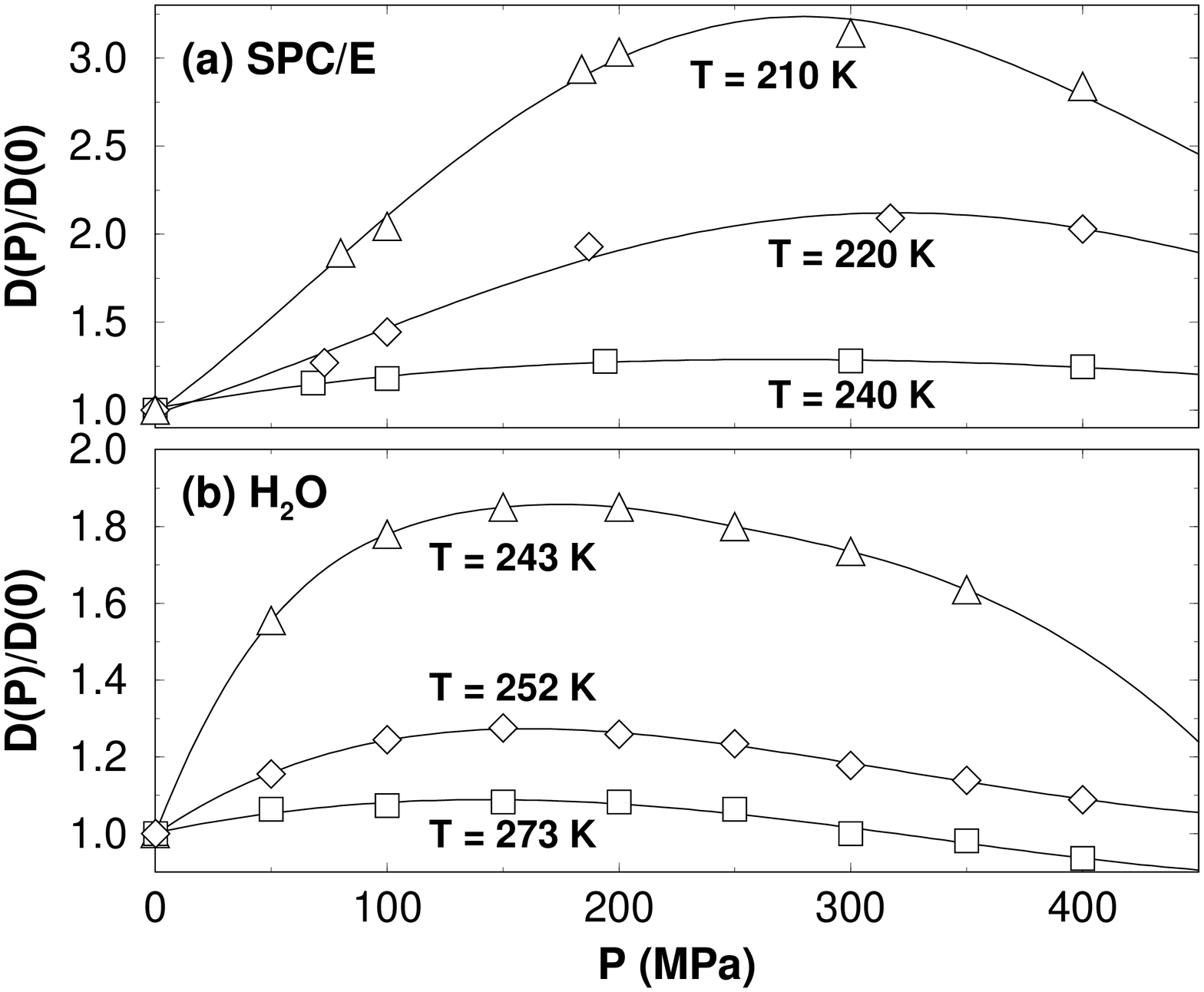,width=3.35in,angle=-90}
\begin{figure*}[htbp]
\begin{center}
\leavevmode
\centerline{\box\figa}
\narrowtext
\caption{$D$ as a function of pressure for various temperatures from (a)
our simulations and (b) NMR studies of water~\protect\cite{prielmeier}.}
\label{fig:DofP}
\end{center}
\end{figure*}
\noindent isobar, our estimates of $D$ are smaller than those
calculated for SPC/E in Ref.~\cite{bc94}, perhaps because
Ref.~\cite{bc94} chooses a different truncation of the electrostatic
terms---highlighting the extreme sensitivity of the dynamics to changes
in the potential.

We fit $D$ by the power law of Eq.~(\ref{eq:D-plaw}) along both
isochores and the estimated isobars for $T \le 300$~K
[Figs.~\ref{fig:D-isochore} and \ref{fig:D-isobar}].  The values of the
two fit parameters $T_c$ and $\gamma$ are given in
Table~\ref{table:plaw-fits}~\cite{vft-note}.  We also include the data
from Ref.~\cite{sns} along the $\rho = 1.0$~g/cm$^3$ isochore and from
Ref.~\cite{francesco} along the $P = -80$~MPa isobar to improve the
quality of the fits.  At $\rho = 1.40$~g/cm$^3$, we exclude $T=210$~K
when fitting $D$ and obtain $T_c = 209.3$, since we expect the power law
of Eq.~(\ref{eq:D-plaw}) to fail for $T \lesssim T_c+5$~K, because
activated processes -- such as ``hopping'' not accounted for in the
idealized MCT -- become significant and aid diffusion.  To demonstrate
the presence of hopping at $\rho=1.40$ and $T=210$~K, we plot the
``self'' part of the van Hove correlation function $G_s(r, t)$, which
measures the distribution of particle displacements $r$ at time $t$, for
several densities at $T=210$~K [Fig.~\ref{fig:vanhove}].  For these
densities where a power law adequately describes $D$, there is a single
peak.  At $\rho = 1.40$~g/cm$^3$, we see a ``shoulder'' in $G_s(r,t)$ at
$r \approx 0.2$~nm in addition to a well-defined peak at $r\approx
0.05$~nm, indicating that particle hopping is significant.

Along the $\rho = 1.00$~g/cm$^3$ we have simulated to significantly
lower $T$, allowing us to study the temperature dependence of $D$ for $T
\lesssim T_c$.  Fig.~\ref{fig:D-arrhenius} shows that the lowest
temperatures are consistent with the Arrhenius form
\begin{equation}
D = D_\infty \exp{(-E/k_BT)}.
\label{eq:arrhenius}
\end{equation}
Arrhenius temperature dependence of $D$ is not surprising in this
region, since for $T<T_c$ it is expected that the energy barriers the
system must overcome to \linebreak

\newbox\figa
\setbox\figa=\psfig{figure=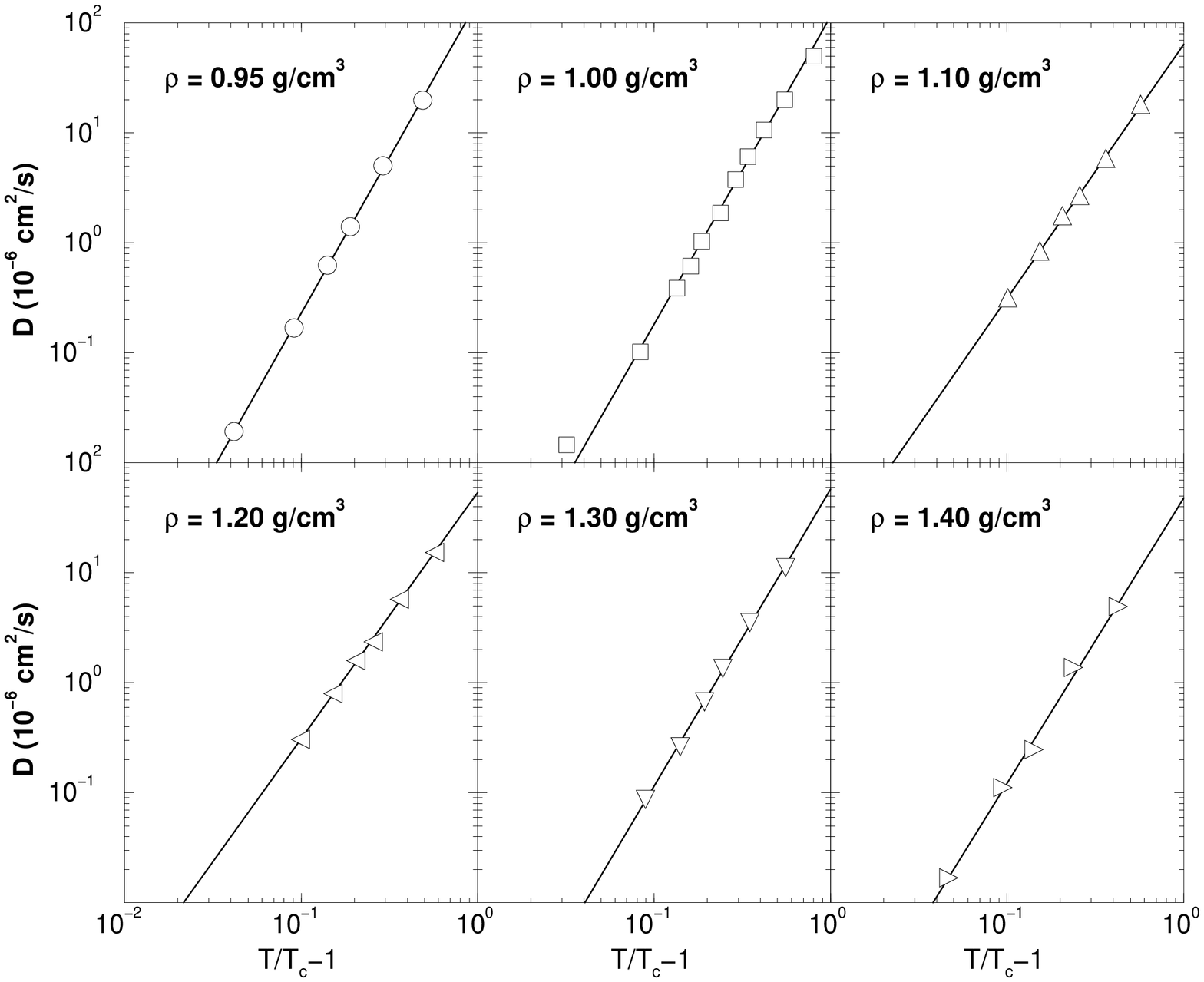,width=3.35in,angle=-90}
\begin{figure*}[htbp]
\begin{center}
\leavevmode
\centerline{\box\figa}
\narrowtext
\caption{Fit of each isochore to the power law $D \sim (T/T_c-1)^\gamma$
predicted by MCT.  We include the data of Ref.~\protect\cite{sns} along
the $\rho = 1.0$~g/cm$^3$ isochore.}
\label{fig:D-isochore}
\end{center}
\end{figure*}
\newbox\figa
\setbox\figa=\psfig{figure=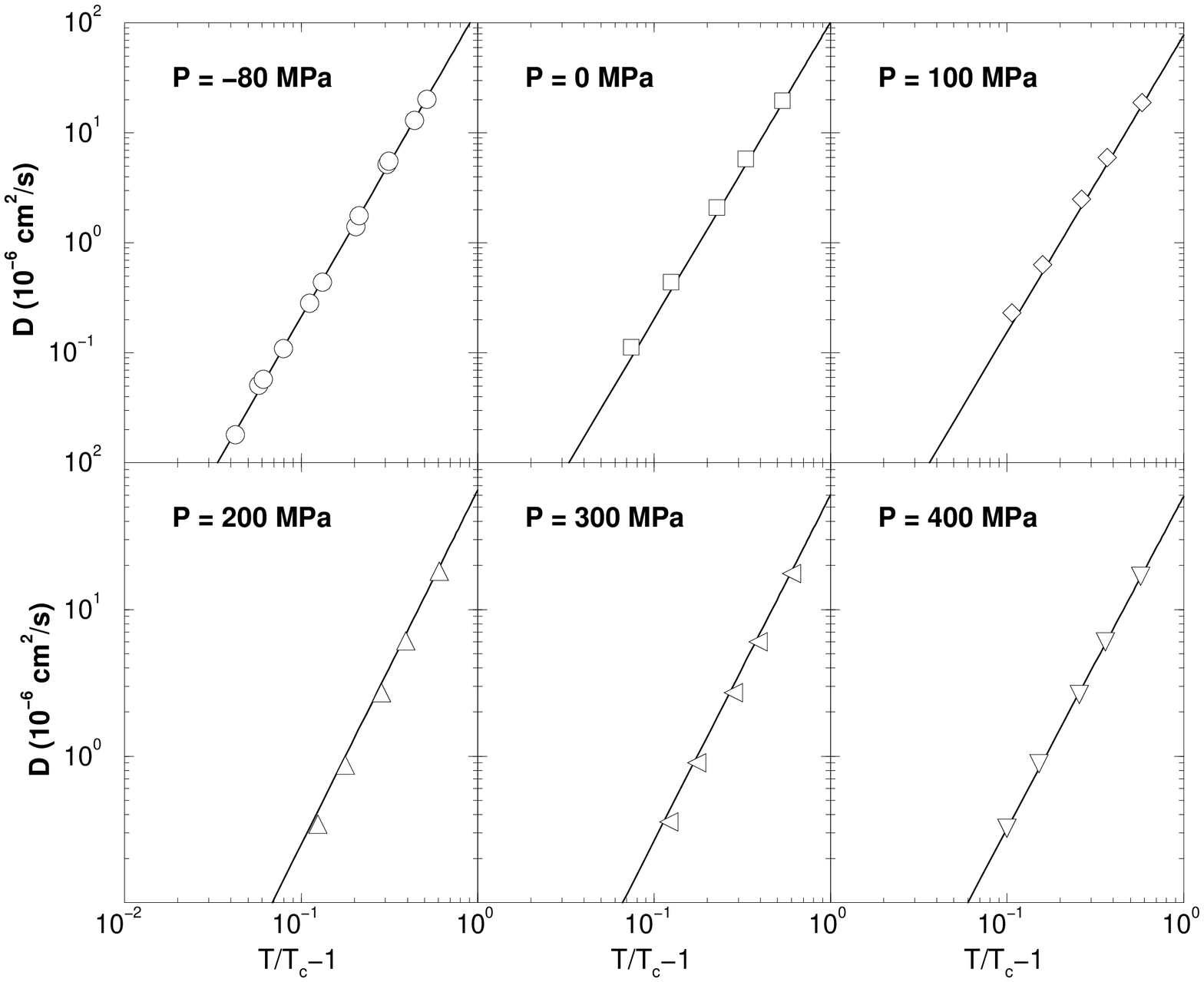,width=3.35in,angle=-90}
\begin{figure*}[htbp]
\begin{center}
\leavevmode
\centerline{\box\figa}
\narrowtext
\caption{Fit of the diffusion constant along each isobar to the power
law $D \sim (T/T_c-1)^\gamma$ predicted by MCT.  We include the data
Ref.~\protect\cite{francesco} along the -80~MPa isobar.}
\label{fig:D-isobar}
\end{center}
\end{figure*}

\noindent rearrange exceed the thermal energy~\cite{st98}.  Hence the
motion of the system is dominated by activated jumps over the energy
barriers, as described by Goldstein~\cite{goldstein}.  We obtain an
activation energy of $E \approx 65$~kJ/mol and extrapolate a glass
transition temperature $T_g \approx 125$~K~\cite{Tg-note}, surprisingly
close to the experimental value of 136~K.  The extrapolated value of
$T_g$ is similar to that estimated in Ref.~\cite{sns} which studied
hydrogen bond dynamics.  Moreover, our results are consistent with a
crossover from ``fragile'' behavior (the behavior described by MCT) for
$T \gtrsim T_c$, to ``strong'' behavior (Arrhenius behavior with
\linebreak 

\newbox\figa
\setbox\figa=\psfig{figure=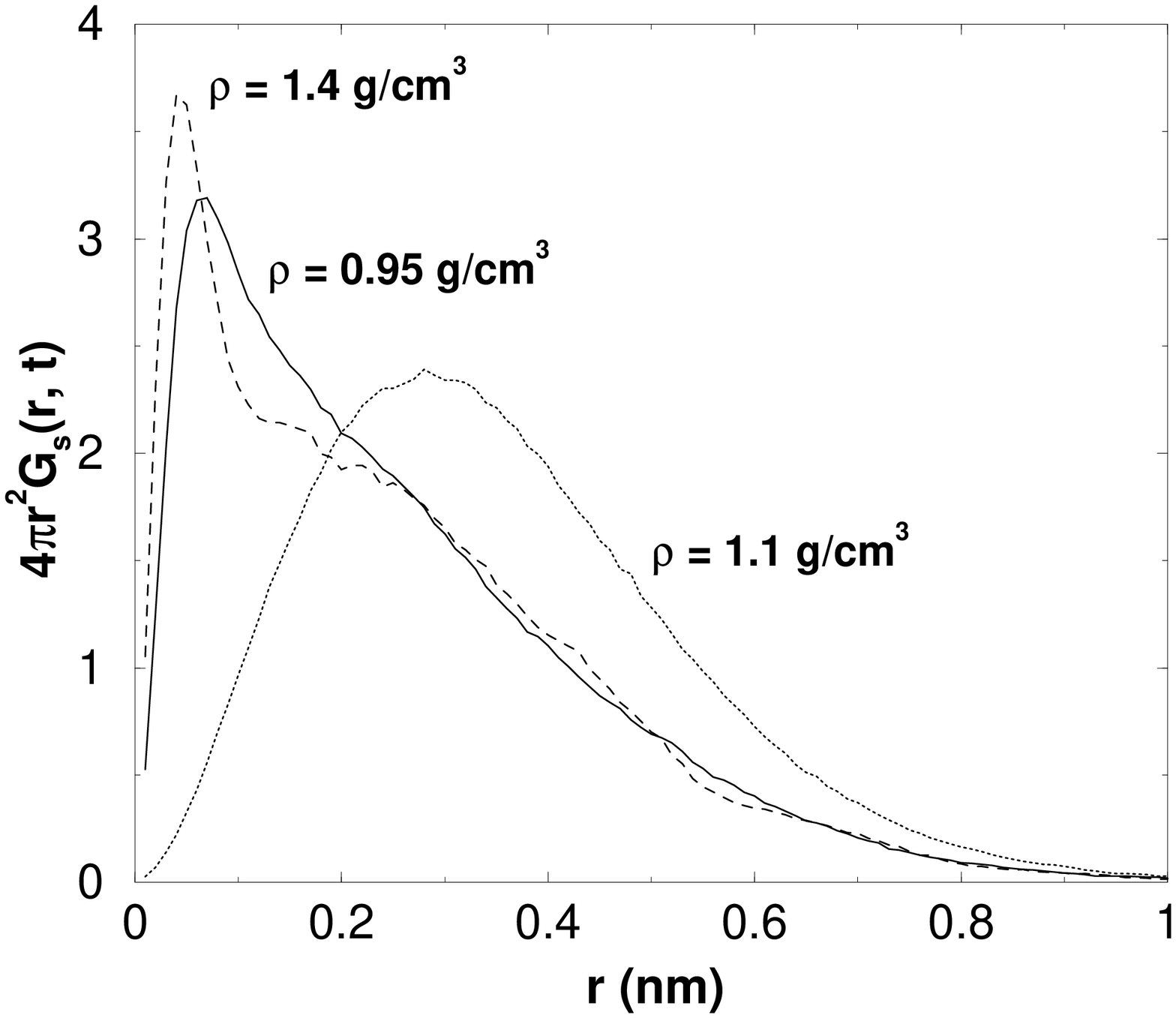,width=3.35in,angle=-90}
\begin{figure}[htbp]
\begin{center}
\leavevmode
\centerline{\box\figa}
\narrowtext
\caption{The van Hove correlation function $G_s(r, t)$ for several
densities at $T=210$~K.  For each curve, $t$ is chosen such that
$\langle r^2(t) \rangle \approx 0.1$~nm$^2$, well inside the diffusive
regime (i.e., where $\langle r^2(t) \rangle$ is linear in $t$).  The
presence of a pronounced shoulder in $G_s(r, t)$ for $\rho =
1.4$~g/cm$^3$ indicates that hopping phenomena are significant, and thus
deviations from power-law dependence are expected.}
\label{fig:vanhove}
\end{center}
\end{figure}
\vspace{-5mm}
\newbox\figa
\setbox\figa=\psfig{figure=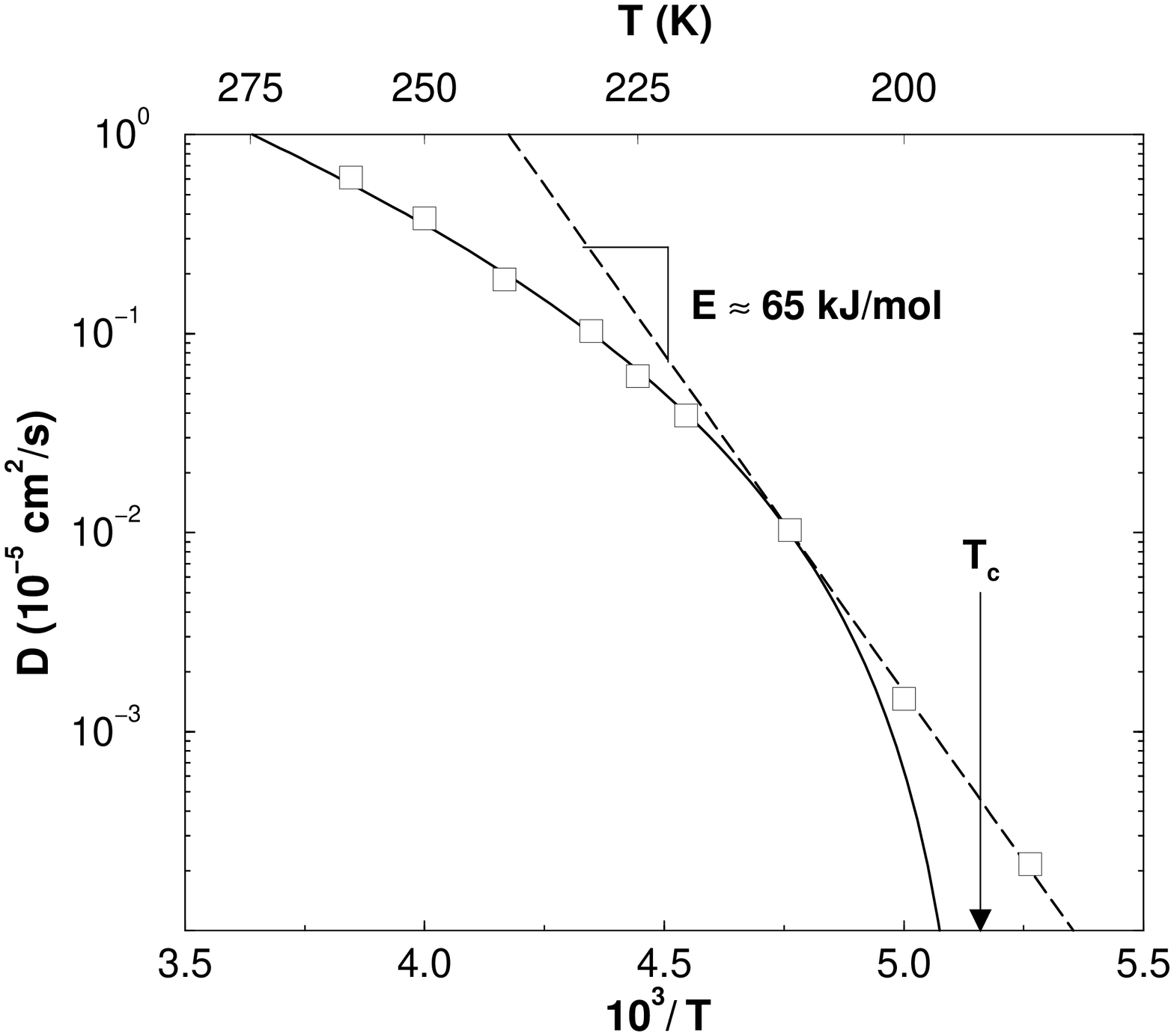,width=3.35in,angle=-90}
\begin{figure}[htbp]
\begin{center}
\leavevmode
\centerline{\box\figa}
\narrowtext
\caption{Arrhenius plot of $D$ shows that the power-law behavior (solid
line) appears to make a smooth crossover to Arrhenius behavior (dashed
line) for $T \protect\lesssim T_c$ with activation energy $E \approx
65$~kJ/mol.  This behavior may be related to a possible
fragile-to-strong transition of the dynamic properties (see discussion
in text).}
\label{fig:D-arrhenius}
\end{center}
\end{figure}
\noindent $E \approx k_BT_g/25 \approx 40$~ kJ/mol for our estimate of
$T_g$) for $T \lesssim T_c$.  The possibility of a ``fragile-to-strong''
crossover in water has been discussed recently based on experimental
findings~\cite{fragile-to-strong}, but lower temperatures are required
to test this possibility in the SPC/E model.

\section{Isochrones of $D$ and the Locus of $T_c(P)$}

To construct isochrones of $D$ (lines of constant $D$), we first
estimate $T(D)$ at values of $D = 10^{-5}$cm$^2$/s,
$10^{-5.5}$~cm$^2$/s, $10^{-6}$cm$^2$/s, and $10^{-7}$cm$^2$/s, using
the fits of Figs.~\ref{fig:D-isochore} and \ref{fig:D-isobar}.  Along
the isobaric paths, we know already $P$ for these points, and along
isochores we may estimate the value of $P$ using the results presented
in Table~\ref{table:state-points}.  We plot the isochrones in
Fig.~\ref{fig:isochrones}.

We also show the the loci of $T_c(P)$ in Fig.~\ref{fig:isochrones}(a),
obtained from the fits in the previous section.  We know $P$ at $T_c$
along the isobaric paths, and we estimate the $P$ at $T_c$ along
isochores by extrapolating $P$ in Table~\ref{table:state-points} to
$T_c$.

\newbox\figa
\setbox\figa=\psfig{figure=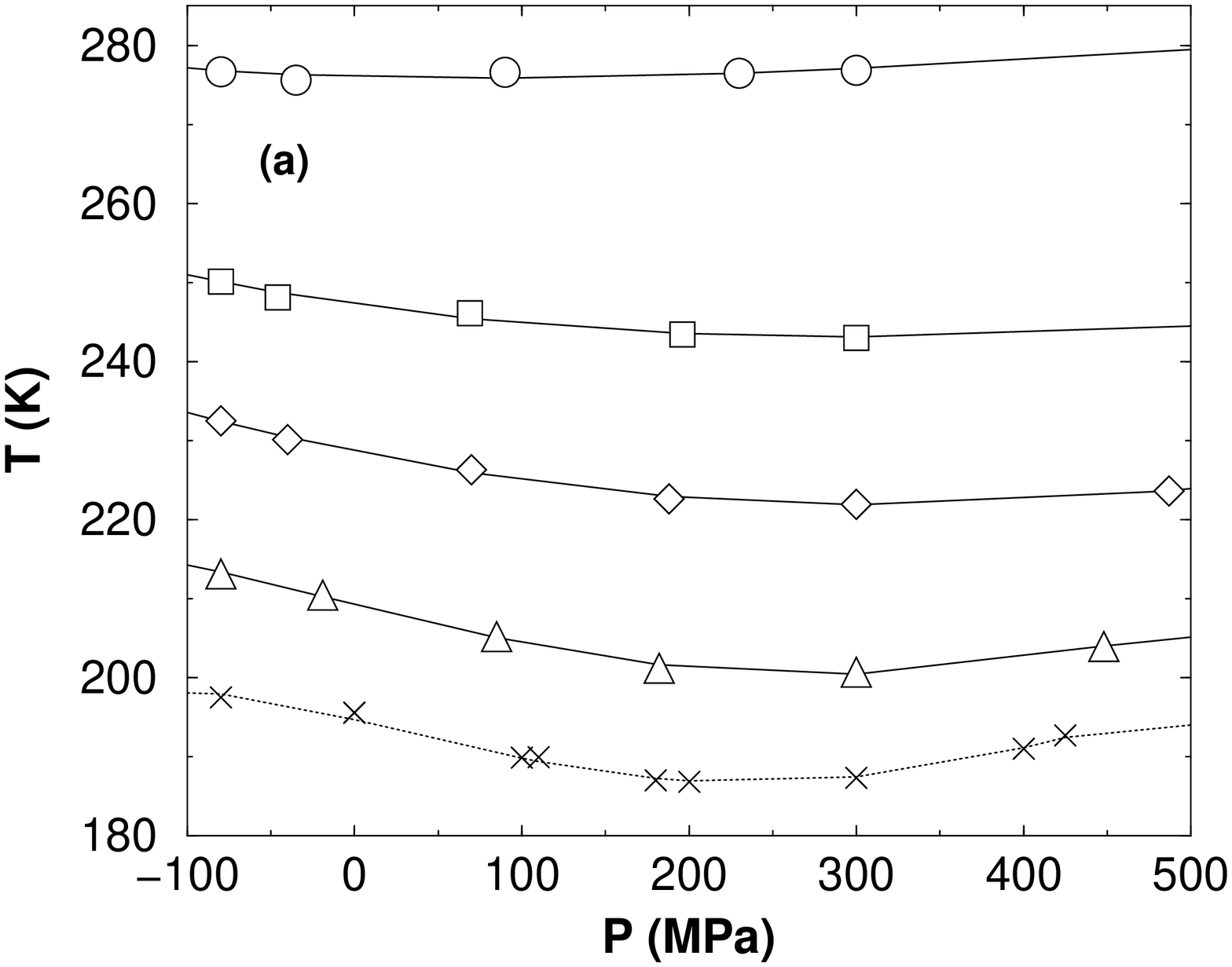,width=3.35in,angle=-90}
\newbox\figb
\setbox\figb=\psfig{figure=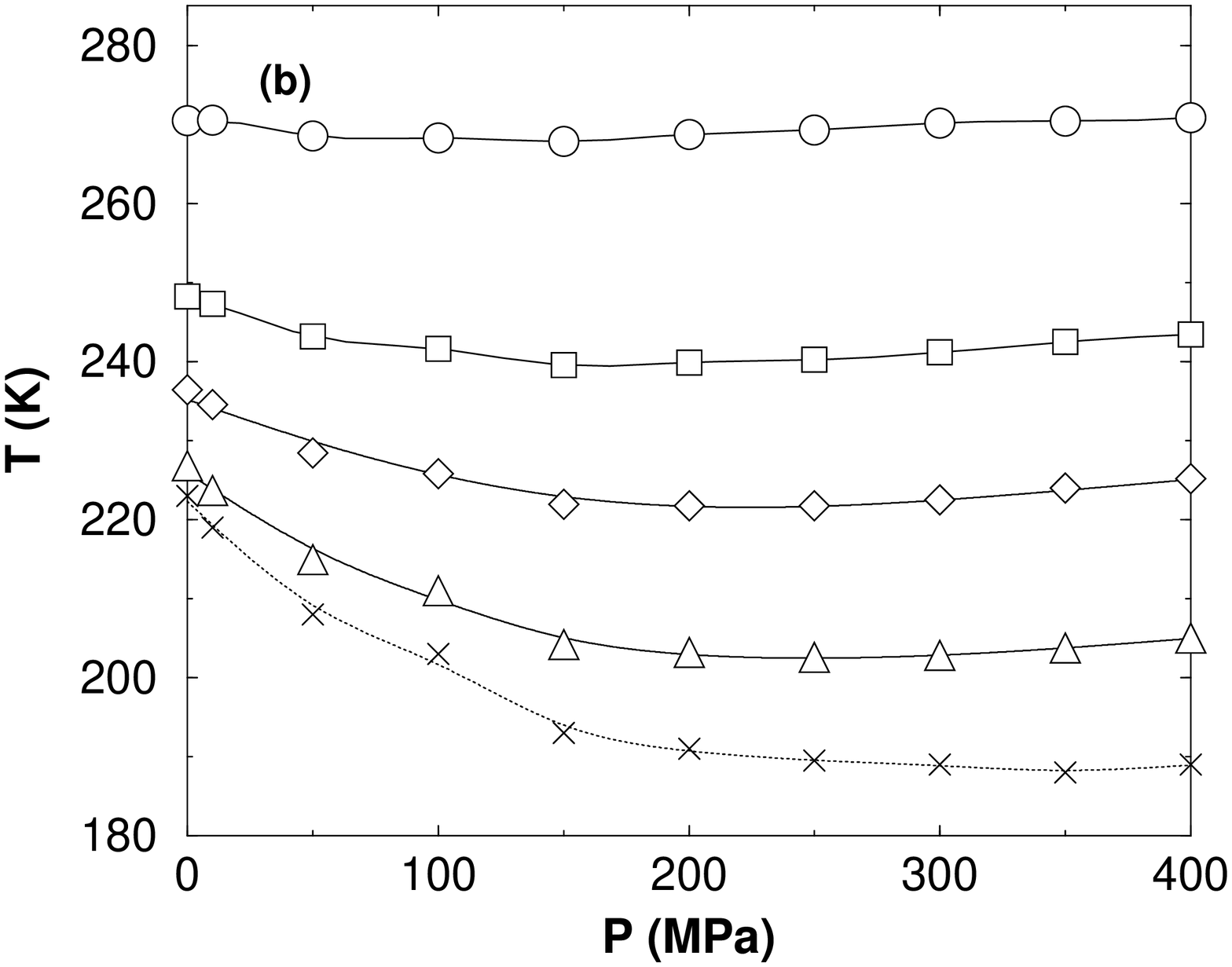,width=3.35in,angle=-90}
\begin{figure*}[htbp]
\begin{center}
\leavevmode
\centerline{\box\figa}
\centerline{\box\figb}
\narrowtext	
\caption{(a) Isochrones of $D$ from simulation.  The lines may be
identified at follows: $D = 10^{-5}$cm$^2$/s ($\circ$); $D =
10^{-5.5}$~cm$^2$/s ($\Box$); $D = 10^{-6}$cm$^2$/s ($\diamond$); $D =
10^{-7}$cm$^2$/s ($\triangle$).  The diffusion is also fit to $D \sim
(T-T_c)^\gamma$.  The locus of $T_c$ is indicated by ($\times$).  (b)
Isochrones of $D$ constructed from the experimental data in
Ref.~\protect\cite{prielmeier}. }
\label{fig:isochrones}
\end{center}
\end{figure*}
\newbox\figa
\setbox\figa=\psfig{figure=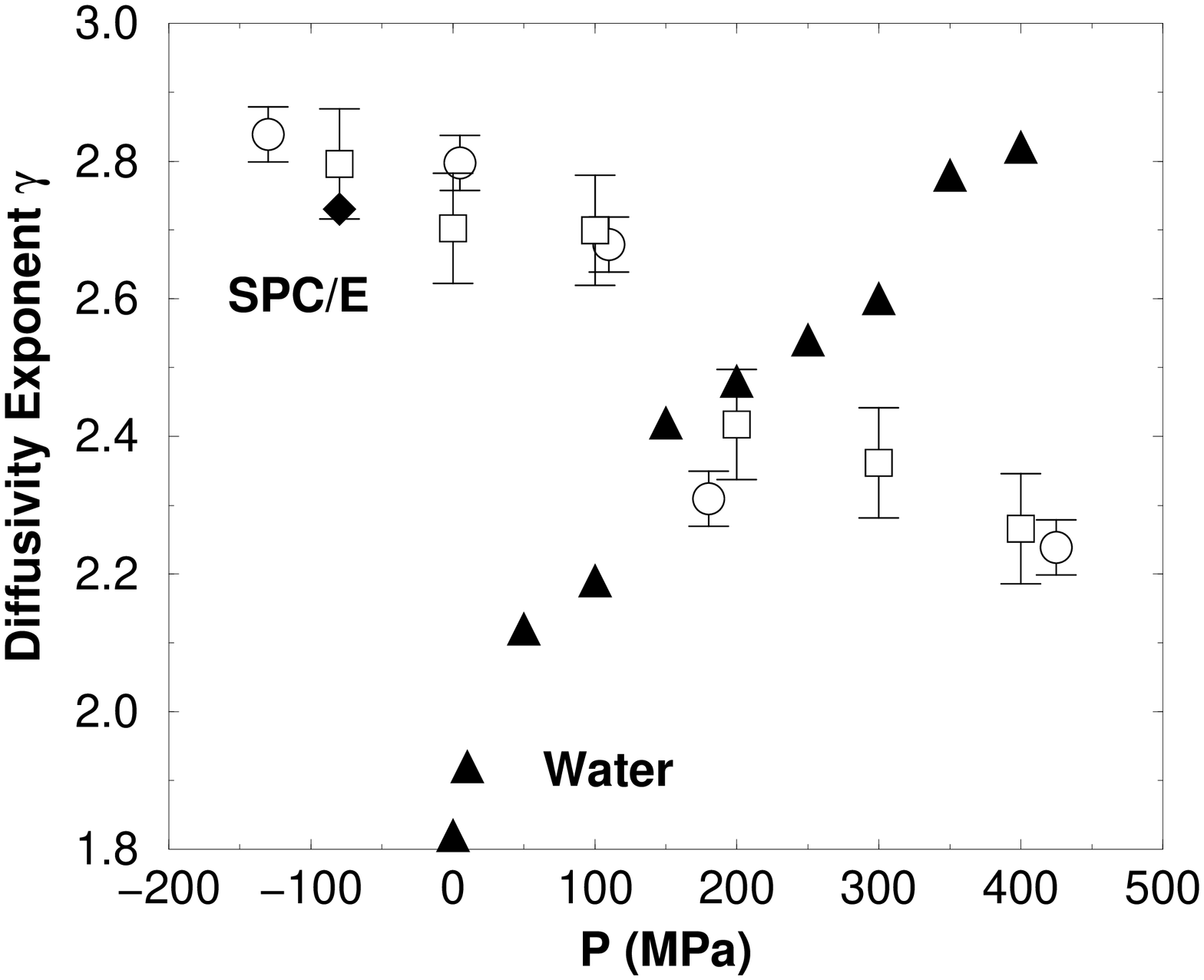,width=3.35in,angle=-90}
\begin{figure*}[htbp]
\begin{center}
\leavevmode
\centerline{\box\figa}
\narrowtext
\caption{Pressure dependence of the diffusivity exponent $\gamma$
defined by $D \sim (T-T_c)^\gamma$.  The symbols are as follows:
($\circ$) $\gamma$ measured from simulation along isochores; ($\Box$)
$\gamma$ measured from simulation along isobaric paths, which are
estimated from the isochoric data; (filled $\Diamond$) $\gamma$ measured
along the -80~MPa isobar in Ref.~\protect\cite{francesco}; (filled
$\triangle$) experimental measurements of $\gamma$ in water from
Ref.~\protect\cite{prielmeier}.  It is clear that the SPC/E potential
fails to reproduce the qualitative behavior of $\gamma$ under pressure
in liquid water.}
\label{fig:gamma}
\end{center}
\end{figure*}

Using the experimental diffusion data of Ref.~\cite{prielmeier}, we also
construct the behavior of the experimental isochrones following the same
technique [Fig.~\ref{fig:isochrones}(b)].  The shape of the locus of
$T_c(P)$ compares well with that observed
experimentally~\cite{prielmeier}, and changes slope at roughly the same
pressure [Fig.~\ref{fig:isochrones}].  Therefore, an explanation of the
SPC/E dynamics using the MCT would support using the MCT framework as an
interpretation of the experimentally found locus of $T_c(P)$.  We find,
however, that $\gamma$ decreases with $P$ for the SPC/E model, while
$\gamma$ increases with $P$ [Fig.~\ref{fig:gamma}].  This disagreement
underscores the need to improve the dynamic properties of water models,
most of which already provide an adequate account of static
properties~\cite{ha98}.

\section{Intermediate Scattering Function}

We plot the intermediate scattering function $F(q_0, t)$ in
Fig.~\ref{fig:isf-allT}(a) for all $T$ along the $\rho = 1.00$~g/cm$^3$
isochore, where $q_0 = 18.55$~nm$^{-1}$, the approximate value of the
first peak of $S(q)$ where the relaxation of $F(q, t)$ is slowest.  We
define the relaxation time $\tau$ by $F(t=\tau) = e^{-1}$.  We show
$\tau$ along isotherms in Fig.~\ref{fig:D-isotherms}(b), from which it
is obvious that $\tau$ has very similar behavior to $D^{-1}$.  Indeed,
MCT predicts that the product $D \tau$ is constant along isochores,
which we test in Fig.~\ref{fig:D-isotherms}(c).  We find that $D \tau$
increases slightly on cooling, but remains relatively constant along
each isochore.  The weak residual \linebreak
\newbox\figa
\setbox\figa=\psfig{figure=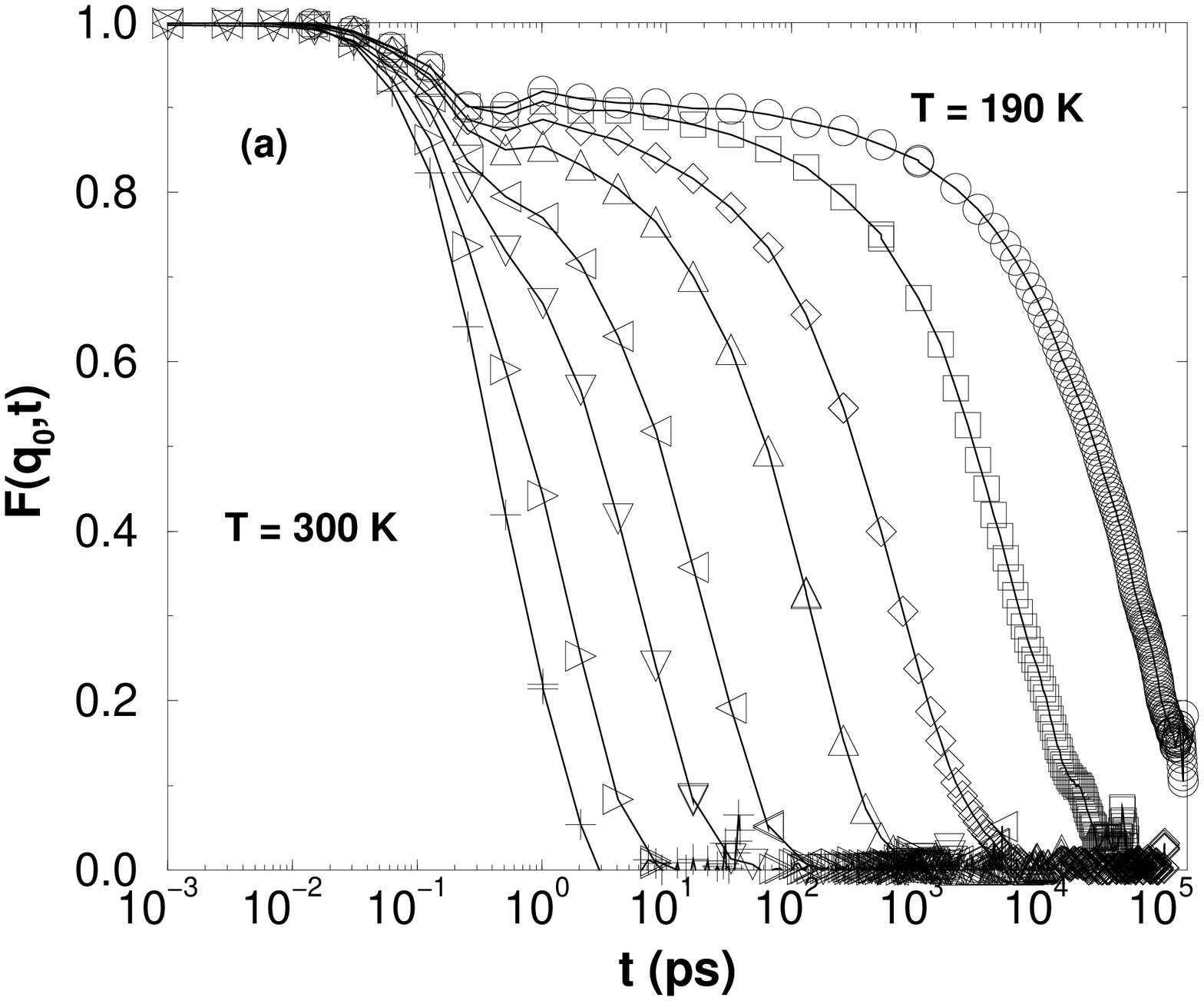,width=3.35in,angle=-90}
\newbox\figb
\setbox\figb=\psfig{figure=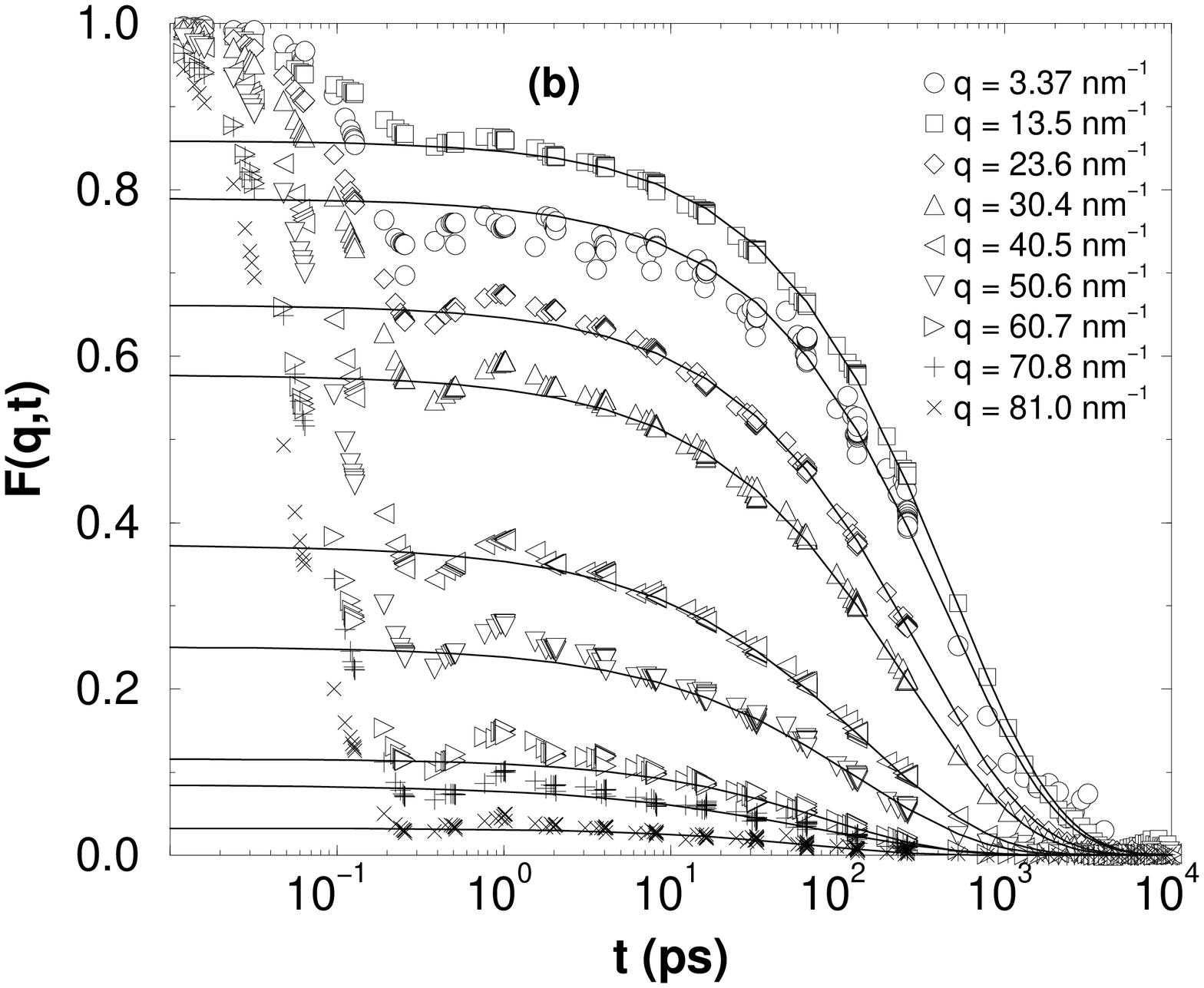,width=3.35in,angle=-90}
\begin{figure*}[htbp]
\begin{center}
\leavevmode
\centerline{\box\figa}
\centerline{\box\figb}
\narrowtext
\caption{The intermediate scattering function $F(q,t)$ (a) for $190 \le
T \le 300$~K along the $\rho = 1.0$~g/cm$^3$ isochore and (b) for many
$q$ values at $T=210$~K and $\rho = 1.00$~g/cm$^3$.  The solid line
shows the fit to Eq.~(\ref{eq:kww}) for $t \ge 2$~ps.  }
\label{fig:isf-allT}
\end{center}
\end{figure*}

\noindent $T$-dependence in $D \tau$ should be subjected to a deeper
scrutiny to find out if it is related to a $q-$vector dependent
correction to scaling (since $D$ is a $q=0$ quantity) or to the
progressive breakdown of the validity of the ideal MCT on approaching
$T_c$.

The study of the time dependence of $F(q_0, t)$ allows us to test the
predicted relation between the exponents $b$ and $\gamma$ (see
Eqs.~(\ref{eq:abg1}) and (\ref{eq:abg2})).  Since the value of $b$ is
completely determined by the value of $\gamma$~\cite{leshouches},
calculation of these exponents for SPC/E determines if MCT is consistent
with our results.  The range of validity of the van Schweidler power law
[Eq.~(\ref{eq:vonSchweidler})] is strongly $q$-dependent~\cite{mayr},
making unambiguous calculation of $b$ difficult.

Fortunately, according to MCT~\cite{fuchs-beta}, at large $q$-vectors,
\linebreak
\newbox\figa
\setbox\figa=\psfig{figure=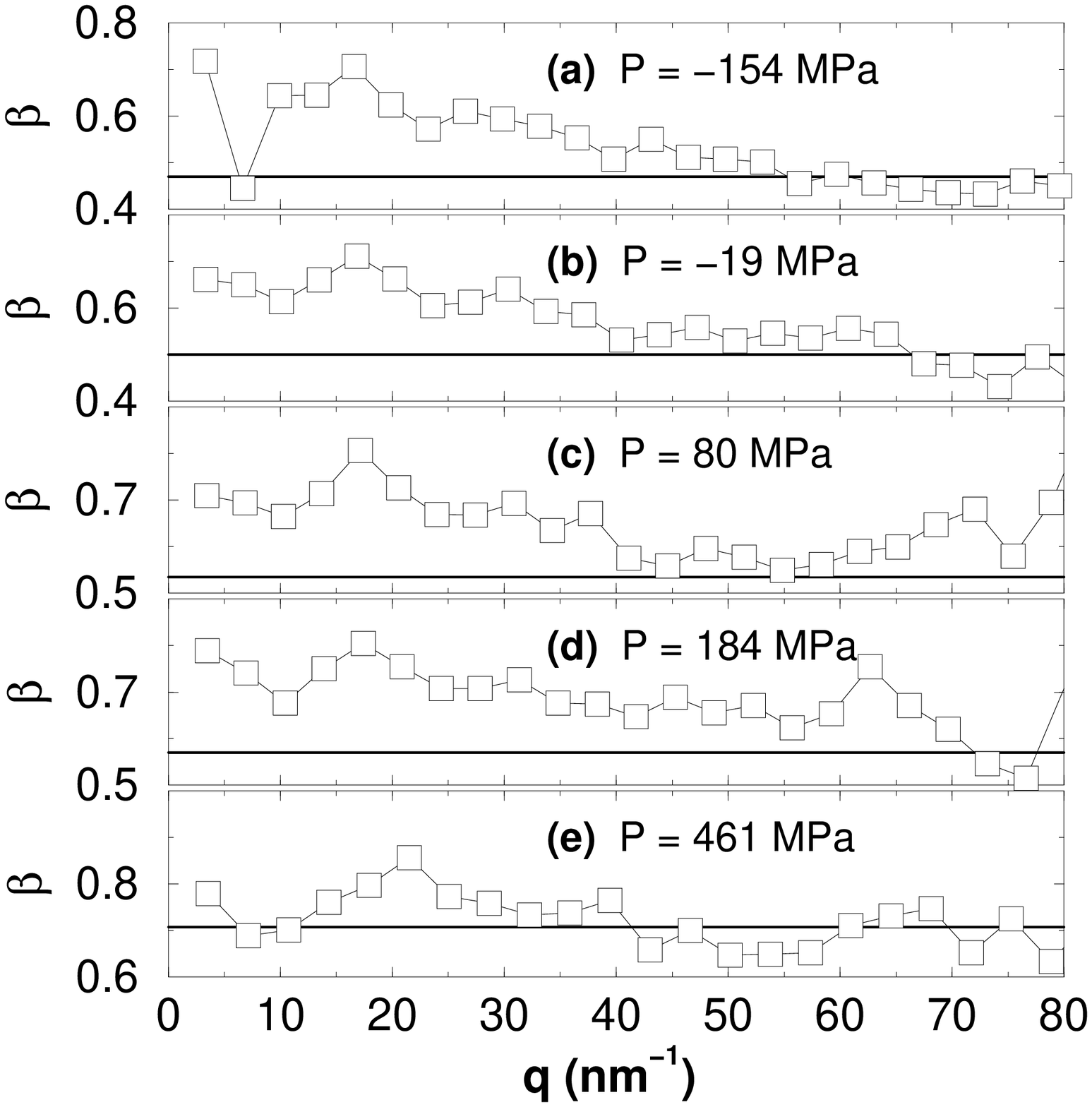,width=3.35in,angle=-90}
\begin{figure*}[htbp]
\begin{center}
\leavevmode
\centerline{\box\figa}
\narrowtext
\caption{Fit of the stretched exponential of Eq.~(\ref{eq:kww}) for $t
\ge 2$~ps at $T=210$~K to both $F_{\mbox{\scriptsize self}}(q,t)$
($\circ$) and $F(q,t)$ ($\Box$) to obtain $\beta$.  The horizontal line
indicates the value predicted by MCT for $b$ using $\gamma$ values
extrapolated from Fig.~\ref{fig:gamma}.  For $P \protect\gtrsim 80$~MPa,
the relaxation of $F(q,t)$ for $q \protect\gtrsim 60$~nm$^{-1}$ comes
almost entirely from the first decay region, so the $\beta$ values
obtained are not reliable in this range. }
\label{fig:betaq}
\end{center}
\end{figure*}

\noindent the stretching exponent $\beta(q)$, which characterizes the
the long-time behavior of $F(q,t)$ (see Eq.~(\ref{eq:kww})), is
controlled by the same exponent $b$ at large $q$.  Fits of $F(q_0, t)$
according to Eq.~(\ref{eq:kww}) are shown for many $q$ values at
$T=210$~K and $\rho = 1.00$~g/cm$^3$.  The same fit quality is observed
for all other low $T$ state points.  The $q$-dependence of $\beta(q)$
for $\rho \le 1.30$~g/cm$^3$ and $T = 210$~K is shown in
Fig.~\ref{fig:betaq}~\cite{betaq-note} for $F(q,t)$.  In addition, we
show the expected value of $b$ according to MCT, using the values of
$\gamma$ extrapolated from Fig.~\ref{fig:gamma}.  The large-$q$ limit of
$\beta$ appears to approach the value predicted by MCT.  Hence we
conclude that the dynamic behavior of the SPC/E potential in the
pressure range we study is consistent with slowing down as described by
MCT [Fig.~\ref{fig:b-gamma}].  We also checked that the values of $b$
calculated from Eq.~(\ref{eq:betaq-limit}) are consistent with the von
Schweidler power law Eq.~(\ref{eq:vonSchweidler}), but that corrections
to scaling in $t^{2b}$ are relevant at several $q$ vectors, as discussed
in Ref.~\cite{francesco}.

\section{Relationship of Structure to Dynamics}
\label{sec:isochrone-dynamics}

The results shown in Fig.~\ref{fig:betaq}, and the observed power-law
dependence of diffusivity, suggest that MCT is able to predict the
dynamical behavior of SPC/E water in a wide range of $P$ and $T$.  As
discussed above, the structure of the liquid changes significantly under
increased pressure.  To highlight the effect of structural changes on
dynamic properties, we consider an approximately isochronic path --
along which $D$ remains nearly constant -- such that the changes in
dynamic properties we observe on increasing $P$ are confined to their
$q$-vector dependence.  We select 5 state points with $D = (0.30 \pm
0.09) \times 10^-6$~cm$^2$/s: (i) $T=220$~K, $\rho = 1.00$~g/cm$^3$,
(ii) $T=210$~K, $\rho = 1.05$~g/cm$^3$, (iii) $T=210$~K, $\rho =
1.10$~g/cm$^3$, (i) $T=210$~K, $\rho = 1.20$~g/cm$^3$, (iv) $T=220$~K,
$\rho = 1.30$~g/cm$^3$, and (v) $T=240$~K, $\rho = 1.40$~g/cm$^3$.  We
show in Fig.~\ref{fig:fits} the $q$-dependence of the
$\alpha$-relaxation time $\tau(q)$ extracted from the fit of $F(q,t)$ to
the stretched exponential of Eq.~(\ref{eq:kww}).  For all state points,
the $q$-dependence of $\tau$ follows the $q$-dependence of $S(q)$, as
commonly observed in supercooled liquids and in solutions of the full
$q$-vector dependent mode coupling equations.  We also note that
$\tau(q)$ is well described by the relation 
\begin{equation}
\tau(q) \propto S(q)/q^2
\label{eq:degenne}
\end{equation}
(the de Gennes narrowing relation), as shown in the same figure.  The
MCT prediction for the $q$-dependence of $\tau$ is often very close to
the relation~(\ref{eq:degenne}).

\section{Discussion}

We have presented extensive simulations that provide evidence for
interpreting the dynamics of the SPC/E potential in the framework of
MCT.  Our calculations also provide a necessary test of the relation
predicted between the diffusivity exponent $\gamma$ and the von
Schweidler exponent $b$ for a wide range of values $\gamma$ and $b$.
Our results support interpretation of the experimental locus $T_c(P)$ as
the locus of MCT transitions.

We found that on increasing pressure, the values of the exponents become
closer to those for hard-sphere ($\gamma=2.58$ and $b=0.545$) and
Lennard-Jones ($\gamma=2.37$ and $b=0.617$) systems~\cite{hs-lj},
thereby confirming that the hydrogen-bond network is destroyed under
pressure and that the water dynamics become closer to that of normal
liquids, where core repulsion dominates.  A significant result of our
analysis is the demonstration that MCT is able to rationalize the
dynamic behavior of the SPC/E model of water at all pressures.  In doing
so, MCT encompasses both the behavior at low pressures, where the
mobility is essentially controlled by the presence of strong energetic
cages of hydrogen bonds, and at high pressures, where the dynamics are
dominated by excluded volume effects.  We also showed how these
structural changes are reflected in the $q$-dependence of $F(q,t)$.

Our results underscore the need to improve the dynamic properties of
potentials for realistic simulations of \linebreak
\newbox\figb
\setbox\figb=\psfig{figure=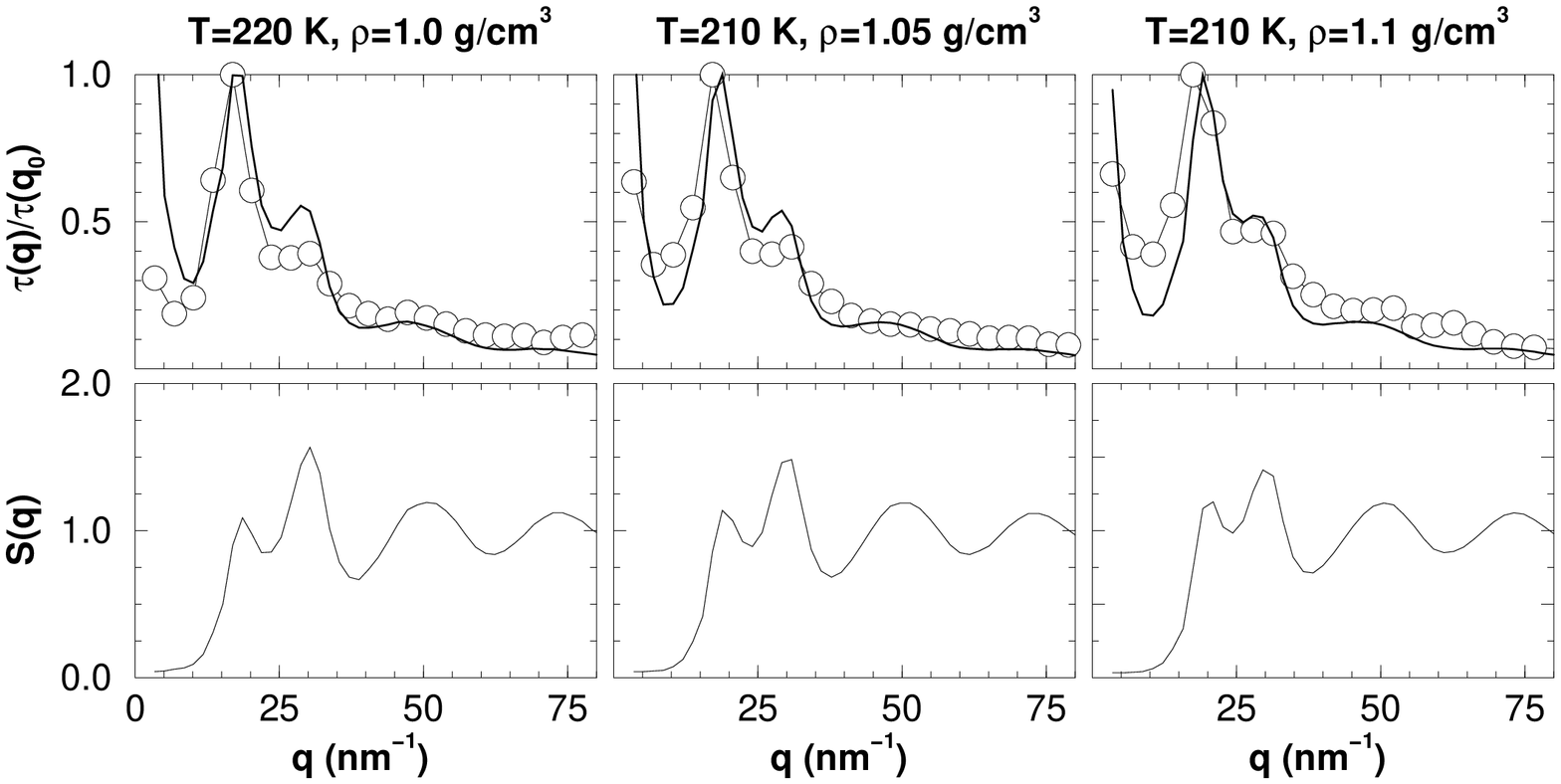,width=3.35in,angle=-90}
\begin{figure*}[htbp]
\begin{center}
\leavevmode
\centerline{\box\figb}
\narrowtext
\begin{minipage}{17.9cm}
\caption{Fitting parameter $\tau(q)$ of Eq.~(\ref{eq:kww}) in relation
to $S(q)$ along approximate isochrones.  The heavy line indicates the
prediction $\tau \sim S(q)/q^2$.}
\end{minipage}
\label{fig:fits}
\end{center}
\end{figure*}

\noindent water and other other materials.  Of the many potentials
available for studying water, only the SPC/E potential is known to
display the power-law dependence of dynamic properties, but even SPC/E
fails to reproduce the power law quantitatively.  A recent study of the
ST2 potential~\cite{st2} found that the $T$-dependence of $D$ is
consistent with an Arrhenius $T$-dependence for $T \gtrsim 300$~K,
crossing over to a another region of Arrhenius behavior for $T \lesssim
275$~K~\cite{paschek-geiger}, in contrast to the non-Arrhenius behavior
observed in real water and to our interpretation based on MCT for $T\ge
T_c$.  The presence of a low-$T$ Arrhenius regime in the ST2 potential
might be due to activated processes, that are expected to dominate the
dynamics of fragile liquids below $T_c$, as we observed for the SPC/E
potential.  Hence the ST2 potential may provide an excellent opportunity
to study these activated processes on a smaller time scale than is
typically observed for most fragile liquids.

Finally, we stress that a full comparison between theory and simulation
data requires a complete solution of the recently proposed molecular-MCT
(the extension of MCT to molecules of arbitrary
shape)~\cite{molecular-mct}.  A detailed solution of the complicated
molecular-MCT equations in such large region of $T$ and $P$ values would
requires computational effort beyond the present possibilities, but a
detailed comparison between molecular-MCT and MD data for one selected
isobar is underway~\cite{fs-inprogress}.

\section{Acknowledgments}
We thank C.A.~Angell, A.~Geiger, E.~La Nave, A.~Rinaldi, S.~Sastry,
A.~Scala and R.J.~Speedy for enlightening discussions and comments on
the manuscript.  We especially thank S. Harrington for his contributions
to the early stages of this work.  We thank the Boston University Center
for Computational Science for access to the 192 processor SGI/Cray
Origin supercomputer.  F.S. is supported in part by MURST (PRIN 98).
The Center for Polymer Studies is supported by the NSF Grant
No. CH9728854.

\newbox\figc
\setbox\figc=\psfig{figure=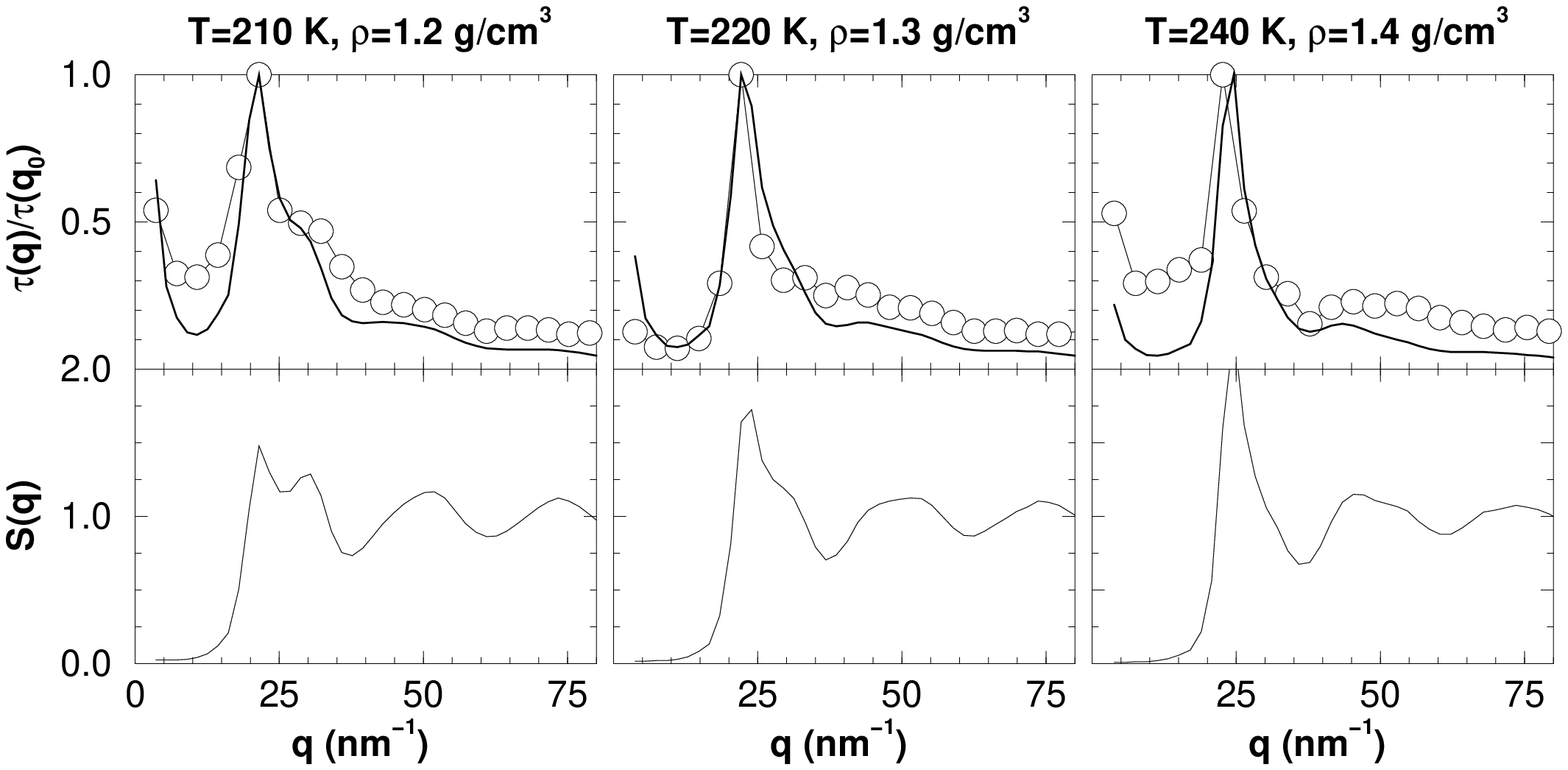,width=3.35in,angle=-90}
\begin{figure*}[htbp]
\begin{center}
\leavevmode
\centerline{\box\figc}
\narrowtext
\end{center}
\end{figure*}

\appendix
\section{Finite-Size Effects} 

Some recent work~\cite{dynamic-het,otp} indicates that significant
finite-size effects can affect results at temperatures close to the MCT
$T_c$.  Most of our systems are farther than 5\% from $T_c$ (i.e.,
$T/T_c-1 \lesssim 0.05$) so that the relatively small size of our system
should not affect our results.  We simulated two independent systems of
1728 molecules at both $T=200$~K and $190~K$ and $\rho = 1.00$~g/cm$^3$
to check if significant finite size effects appear at low $T$.  The
results are shown in Table.~\ref{table:1728mols}.  We observe no
significant deviations from the system of 216 molecules at $T=200$~K.
Hence we believe that no strong finite-size effects are present in the
216 molecule system for $T\ge200$~K.  However at $T=190$~K, the
potential energy of the larger system appears to be significantly
smaller than that in the smaller system.  We lack adequate computer
resources to make a reliable estimate the diffusivity in the larger
system, but simulations are continuing in order to check the possible
finite-size effects at this temperature.

\end{multicols}

\noindent\begin{minipage}{17.9cm}
\begin{table}
\caption{Summary of the state points simulated with 216 molecules
interacting via the SPC/E potential.  For all state points, the
uncertainty in the potential energy $U$ is less than 0.05~kJ/mol.  The
uncertainty in the diffusion constant $D$ is approximately $\pm 4$, in
the last digit shown.  State points we equilibrated for a time
$t_{\mbox{\scriptsize eq}}$, followed by ``data collection'' runs of
duration $t_{\mbox{\scriptsize data}}$.}
\medskip
\begin{tabular}{c|c|ccc|cc}
T & $\rho$ (g/cm$^3$) & U (kJ/mol) & P (MPa) & D ($10^{-6}$ cm$^2$/s) &
$t_{\mbox{\scriptsize eq}}$ (ns) & $t_{\mbox{\scriptsize data}}$ (ns) \\
\tableline
190 & 1.00 & $-55.00 $ & $5 \pm 20$ & 0.00022  & 60 & 140 \\
\tableline
200 & 1.00 & $-54.44 $ & $-1 \pm 18$ & 0.0015 & 30 & 100 \\
\tableline
210 & 0.90 & $-53.45 $ & $-298 \pm 15$ & 0.0292 & 8 & 100 \\
    & 0.95 & $-53.84 $ & $-154 \pm 9$ & 0.0193 & 25 & 100\\
    & 1.00 & $-53.70 $ & $-19 \pm  11$ & 0.103 & 35 & 100\\
    & 1.05 & $-53.43 $ & $80  \pm  12$ & 0.227 & 30 & 80\\
    & 1.10 & $-53.24 $ & $184 \pm  13$ & 0.317 & 30 & 80\\
    & 1.20 & $-53.13 $ & $461  \pm 14$ & 0.304 & 25 & 80\\
    & 1.30 & $-53.20$ & $901  \pm 15$ & 0.0871 & 25 & 100\\
    & 1.40 & $-52.98$ & $1624 \pm 14$ & 0.00486 & 30 & 100\\
\tableline
220 & 0.95 & $-53.00$  & $-150 \pm 6$ & 0.168 & 15 & 15 \\
    & 1.00 & $-52.87$  & $-21 \pm 10$ & 0.389 & 15 & 15 \\
    & 1.05 & $-52.73$  & $73 \pm 8$  & 0.558 & 15 & 15 \\
    & 1.10 & $-52.59$  & $187 \pm 8$ & 0.847 & 15 & 15\\
    & 1.15 & $-52.53$ & $317 \pm 8$ & 0.918 & 2 & 15\\
    & 1.20 & $-52.48$  & $480 \pm 9$ & 0.801 & 15 & 15\\
    & 1.25 & $-52.47$ & $687 \pm 9$ & 0.594 & 3 & 15\\
    & 1.30 & $-52.49$  & $951 \pm 12$ & 0.263 & 15 & 15\\
    & 1.40 & $-52.57$ & $1670 \pm 15$ & 0.0169 & 18 & 15\\
\tableline
230 & 0.95  & $-52.14 $ & $ -155 \pm 8$ & 0.625 & 4 & 5\\
    & 1.00  & $-52.06 $ & $ -41  \pm 9$ & 1.03  & 4 & 5\\
    & 1.05  & $-52.01 $ & $ 70   \pm 10$ & 1.34  & 4 & 5\\
    & 1.10  & $-51.90 $ & $ 193  \pm 12$ & 1.77  & 4 & 5\\
    & 1.20  & $-51.85 $ & $ 501  \pm 13$ & 1.59  & 4 & 5\\
    & 1.30  & $-51.90 $ & $ 994  \pm 14$ & 0.672 & 4 & 5\\
    & 1.40  & $-51.82 $ & $ 1752 \pm 17$ & 0.112 & 4 & 5\\
\tableline
240 & 0.95  & $-51.33 $ & $-153  \pm 8$ & 1.41 & 7 & 5\\
    & 1.00  & $-51.35 $ & $-45 \pm   9$ & 1.87 & 7 & 5\\
    & 1.05  & $-51.34 $ & $68 \pm    9$ & 2.44 & 7 & 5\\
    & 1.10  & $-51.28 $ & $195 \pm  10$ & 2.70 & 7 & 5\\
    & 1.20  & $-51.24 $ & $527 \pm  11$ & 2.37 & 7 & 5\\
    & 1.30  & $-51.25 $ & $1035 \pm  4$ & 1.35 & 7 & 5\\
    & 1.40  & $-51.24 $ & $1828 \pm 12$ & 0.249 & 12 & 5\\
\tableline
260 & 0.95 & $-49.68 $ & $-148 \pm  9$ & 5.04 & 5 & 3\\
    & 1.00 & $-49.87 $ & $-43  \pm 10$ & 6.08 & 5 & 3\\
    & 1.05 & $-49.93 $ & $77   \pm 11$ & 5.91 & 5 & 3\\
    & 1.10 & $-50.00 $ & $212  \pm 11$ & 5.88 & 5 & 3\\
    & 1.20 & $-50.10 $ & $572  \pm 13$ & 5.74 & 5 & 3\\
    & 1.30 & $-50.14 $ & $1127 \pm 14$ & 3.54 & 5 & 3\\
    & 1.40 & $-49.97$ & $1979 \pm 14$ & 1.39 & 5 & 3\\
\tableline
300 & 0.95 & $-46.80 $ & $-109 \pm 12$ & 19.9 & 0.5 & 1\\
    & 1.00 & $-47.20 $ & $-13 \pm 13 $ & 20.0 & 0.5 & 1\\
    & 1.05 & $-47.49 $ & $112 \pm 14 $ & 18.3 & 0.5 & 1\\
    & 1.10 & $-47.65 $ & $264 \pm 14 $ & 18.2 & 0.5 & 1\\
    & 1.20 & $-47.95 $ & $678 \pm 16 $ & 15.3 & 0.5 & 1\\
    & 1.30 & $-48.06 $ & $1293 \pm 18$ & 11.2 & 0.5 & 1\\
    & 1.40 & $-47.88$ & $2222 \pm 19$ & 4.95 & 0.5 & 1\\
\tableline
350 & 0.90 & $-43.21$ & $-105 \pm 16$ & 61.1 & 0.5 & 40 ps\\
    & 1.00 & $-44.35 $ & $62  \pm  18$ & 49.7 & 0.5 & 40 ps\\
    & 1.10 & $-45.15 $ & $358 \pm  20$ & 38.1 & 0.5 & 40 ps\\
    & 1.20 & $-45.56 $ & $828 \pm  22$ & 27.0 & 0.5 & 40 ps\\
    & 1.30 & $-45.76 $ & $1504 \pm 25$ & 18.0 & 0.5 & 40 ps\\
    & 1.40 & $-45.50$ & $2522 \pm 26$ & 13.9 & 0.5 & 40 ps\\
\end{tabular}
\label{table:state-points}
\end{table}
\end{minipage}

\noindent\begin{minipage}{17.9cm}
\begin{table}
\caption{Fitting parameters to the power law predicted by MCT, for $T
\le 300$~K.  Note that the state points ($\rho = 1.00$~g/cm~$^3$,
$T=200$~K and $T=190$~K) and ($\rho = 1.40$~g/cm~$^3$, $T=210$~K) are
not included in the fit because they are very close to $T_c$, and so do
not conform to the power law.}
\medskip
\begin{tabular}{cccc}
$\rho$ (g/cm$^3$) or $P$ (MPa) & $D_0$ ($10^{-6}$ cm$^2$/s) & 
$T_c (K) $ & $\gamma$ \\
\hline
0.95 & 159  & 201.4 & 2.84 \\
1.00 & 114  & 193.6 & 2.80  \\
1.05 & 78.5 & 188.3 & 2.67 \\
1.10 & 63.9 & 188.6 & 2.31 \\
1.20 & 54.0 & 189.9 & 2.24 \\
1.30 & 57.3 & 192.9 & 2.70 \\
1.40 & 47.8 & 210.2 & 2.59 \\
\hline
-80 &  133  & 197.9 & 2.79 \\
0   &  102  & 193.9 & 2.62 \\
100 &  77.2 & 187.8 & 2.61 \\
200 &  65.1 & 188.8 & 2.50 \\
300 &  60.8 & 188.8 & 2.42 \\
400 &  59.1 & 190.7 & 2.25 \\
\end{tabular}
\label{table:plaw-fits}
\end{table}
\end{minipage}

\noindent\begin{minipage}{17.9cm}
\begin{table}
\caption{Summary of the state points simulated with 1728 molecules to
check for finite-size effects.  For both state points, the uncertainty
in the potential energy $U$ is less the 0.04~kJ/mol.}
\medskip
\begin{tabular}{c|c|ccc|cc}
T & $\rho$ (g/cm$^3$) & U (kJ/mol) & P (MPa) & D ($10^{-6}$ cm$^2$/s) &
$t_{\mbox{\scriptsize eq}}$ (ns) & $t_{\mbox{\scriptsize data}}$ (ns) \\
\tableline
190 & 1.00 & $-55.12 $ & $6 \pm 8$ & ---  & 60 & 40 \\
\tableline
200 & 1.00 & $-54.41 $ & $-4 \pm 7$ & $0.0020 \pm 0.0007$ & 30 & 35 \\
\end{tabular}
\label{table:1728mols}
\end{table}
\end{minipage}


\begin{references}
\bibitem{not-com} In this work we study the motion of oxygen atoms, and
not the center of mass motion.  However, for the SPC/E model, the center
of mass is offset from the oxygen along the H-O-H bisector by only
0.06~\AA\, so the difference is negligible.
\bibitem{Tg-note} We estimate $T_g$ by extrapolating
Eq.~(\ref{eq:arrhenius}) to the temperature where $D$ reaches
$10^{-18}$~cm$^2$/s, a typical value of $D$ at the glass transition.
\bibitem{betaq-note} We do not show the behavior of $\beta(q)$ for $\rho
= 1.40$~g/cm$^3$ and $T = 210$~K since the MCT predictions are not
expected be valid for this state point, as discussed in
Sec.~\ref{sec:diffusion}.
\end{references}
\end{document}